\documentclass{aa}

\usepackage[colorlinks=true,allcolors=blue]{hyperref}
\usepackage[svgnames]{xcolor}
\usepackage{graphicx}
\usepackage{amsmath}
\usepackage{amssymb}
\usepackage{orcidlink}
\usepackage{array}
\usepackage{arydshln, booktabs}
\usepackage{tabularx}
\usepackage{cellspace}
\usepackage{bm}
\usepackage{upgreek}
\usepackage{txfonts}
\usepackage{lscape}
\usepackage{tikz}

\defcitealias{Rihtarsic2026}{GR26}
\defcitealias{Richard2021}{JR21}
\defcitealias{Motta2018}{M18}

\renewcommand{\arraystretch}{1.05}
\definecolor{mycyan}{HTML}{00BFFF}
\definecolor{mygold}{HTML}{FFC800}

\newcommand{\cyancircle}[1]{%
  \tikz[baseline=(c.base)]{
    \node[draw=mycyan,
          circle,
          inner sep=1pt,
          minimum size=9pt,
          line width=0.75pt] (c) {\small #1};}}

\newcommand{\goldcircle}[1]{%
\hspace{1pt}%
  \tikz[baseline=(c.base)]{
    \node[draw=mygold,
          circle,
          inner sep=1pt,
          minimum size=9pt,
          line width=0.75pt] (c) {\small #1};}}

\begin{document}

   \title{An ALMA sub-kiloparsec test of the Kennicutt--Schmidt relation in~a~strongly lensed dusty star-forming galaxy at $z\sim2.78$}
   
   \author{
   Carla Cornil-Ba\"{i}otto 
      \inst{\textcolor{blue}{1},\textcolor{blue}{2}}
      \fnmsep\thanks{carla.cornil@postgrado.uv.cl}\orcidlink{0009-0008-4494-1642},
   Juan Molina 
      \inst{\textcolor{blue}{1},\textcolor{blue}{2}}\orcidlink{0000-0002-8136-8127}, 
   Guillaume Mahler 
      \inst{\textcolor{blue}{3}}\orcidlink{0000-0003-3266-2001}, 
   Gregor Rihtar\v{s}i\v{c} 
      \inst{\textcolor{blue}{4}}\orcidlink{0009-0009-4388-898X}, \\
   M\'{e}d\'{e}ric Boquien 
      \inst{\textcolor{blue}{5}}\orcidlink{0000-0003-0946-6176}, 
   Maru\v{s}a Brada\v{c} 
      \inst{\textcolor{blue}{4}}\orcidlink{0000-0001-5984-0395}, 
   Eduardo Ibar 
      \inst{\textcolor{blue}{1},\textcolor{blue}{2}}\orcidlink{0009-0008-9801-2224}, 
   Anthony H.~Gonzalez 
      \inst{\textcolor{blue}{6}}\orcidlink{0000-0002-0933-8601}, \\
   Tom\'{a}s Verdugo 
      \inst{\textcolor{blue}{7}}\orcidlink{0000-0003-4062-6123}, 
   Omar L\'{o}pez-Cruz 
      \inst{\textcolor{blue}{8}}\orcidlink{0000-0002-1381-7437}, 
   Juan Maga\~{n}a 
      \inst{\textcolor{blue}{9}}\orcidlink{0000-0003-1750-4769},
   \and Ver\'{o}nica Motta 
      \inst{\textcolor{blue}{1}}\orcidlink{0000-0003-4446-7465} 
   }
   \institute{Instituto de F\'{i}sica y Astronom\'{i}a, Universidad de Valpara\'{i}so, Avda.~Gran Breta\~{n}a 1111, Valpara\'{i}so, Chile
         \and
              Millenium Nucleus for Galaxies (MINGAL)
         \and  
              STAR Institute, Quartier Agora - All\'{e}e du six Août, 19c B-4000 Li\`{e}ge, Belgium
         \and 
              Faculty of Mathematics and Physics, Jadranska ulica 19, SI-1000 Ljubljana, Slovenia
        \and
              Universit\'{e} C\^{o}te d'Azur, Observatoire de la C\^{o}te d'Azur, CNRS, Laboratoire Lagrange, 06000 Nice, France 
        \and
              Department of Astronomy, University of Florida, Bryant Space Science Center, Gainesville, FL 32611, USA
        \and
              Universidad Nacional Aut\'{o}noma de M\'{e}xico, Instituto de Astronom\'{i}a, A.P.\:106, 22800, Ensenada, B.C., M\'{e}xico
        \and  
              Instituto Nacional de Astrof\'{i}sica \'{O}ptica y Electr\'{o}nica (INAOE), Luis E.\:Erro No.\:1, C.P. 72840 Puebla, M\'{e}xico
        \and
              Escuela de Ingenier\'{i}a, Universidad Central de Chile, Av.\:Francisco de Aguirre 0405, 171-0164 La Serena, Coquimbo, Chile
        }

   \date{Received 09/07/2026; accepted 28/10/2026}

   \abstract
   {Star formation is a key process driving galaxy evolution, but understanding how it proceeds within the interstellar medium (ISM) requires observations that resolve the cold molecular gas and dust on sub-kiloparsec (sub-kpc) scales. At $z$\,$\sim$\,$2$--$3$, near the peak of cosmic star formation, such spatial resolution can generally only be achieved through strong gravitational lensing.}
   {We present an ALMA study of SMM\:J065837.6$-$555705, hereafter SMM\,J0658, a $20\times$-magnified main-sequence star-forming galaxy at $z_{\rm CO}$\,=\,$2.7768$\,$\pm$\,$0.0002$, lensed by the Bullet Cluster (1E\,0657$-$56; $z$\,=\,$0.296$) into three images. We use 0\farcs2 and 0\farcs6 ALMA Band-3 data to test the Kennicutt--Schmidt (KS) relation at sub-kpc scales at cosmic noon.}
   {Using a JWST-based parametric strong-lensing model of the Bullet Cluster, we reconstruct source-plane \text{CO(3--2)} and rest-frame 803~$\upmu$m dust-continuum maps down to physical scales of $\sim$\,$200$~pc. We derive spatially resolved maps of the intrinsic molecular gas mass surface density ($\Sigma_{\rm mol}$), star formation rate surface density ($\Sigma_{\rm SFR}$), molecular gas depletion time ($\tau_{\rm dep}$), and quantify the scale dependence of the CO-to-dust flux ratio over apertures from 200~pc to 3.2~kpc across the reconstructed source.}
   {The 0\farcs2 ALMA data reveal massive ($M_{\rm mol}$\,$\sim$\,$10^9\pm0.3~{\rm M}_\odot$) and dense star-forming clumps, with $\Sigma_{\rm mol}$\,$\sim$\,1.3--2.3~$\times10^3~{\rm M}_\odot\,{\rm pc}^{-2}$, $\Sigma_{\rm SFR}$\,$\sim$\,0.5--2.3~${\rm M}_\odot\,{\rm yr}^{-1}\,{\rm kpc}^{-2}$, and $\tau_{\rm dep}$\,$\sim$\,0.82--2.56~Gyr. The observed CO--dust spatial decorrelation produces a strong scale dependence in $\tau_{\rm dep}$. We find a KS-relation breakdown scale of 0.8\,$\pm$\,0.1~kpc in SMM\,J0658, consistent with the resolution of individual regions tracing distinct evolutionary stages of the star formation cycle within the ISM. The ALMA and JWST data also reveal evidence for a lensed galaxy pair at $z$\,$\sim$\,$2.78$, although it remains uncertain whether SMM\,J0658 is undergoing an early-stage interaction that could contribute to an enhanced star-formation efficiency in localized regions of the ISM.}
   {Our results show that the apparent universality of the KS relation breaks down once individual regions of the star-formation process are spatially resolved, in line with observations from the local universe to $z$\,$\sim$\,$1$. We extend this picture to $z$\,$\sim$\,$2.78$, opening a new window onto the resolved properties of the clumpy cold ISM and the star formation cycle at cosmic noon.}
   

   \titlerunning{Sub-kiloparsec test of the Kennicutt--Schmidt relation at $z\sim2.78$}
   \authorrunning{Cornil-Ba\"{i}otto et al.}
   \maketitle

\section{Introduction}

   Mapping the cold molecular gas and dust at sub-kiloparsec (sub-kpc) scales in galaxies is essential for understanding the physical processes driving star formation in the interstellar medium (ISM), particularly at $z$\,$\sim$\,$2$--$3$, during the peak epoch of cosmic star formation \citep[cosmic noon;][]{Madau2014}.

   Galaxies at cosmic noon exhibit significantly higher star formation rates (SFRs) and molecular gas fractions than galaxies in the local universe \citep[e.g.,][]{Tacconi2010, Tacconi2013}, often leading to unstable, clumpy, and turbulent rotating disks \citep[e.g.,][]{Forster2009, Genzel2011, Romeo2014, Noble2017}. This elevated star formation activity is closely linked to large cold molecular gas reservoirs, which provide the primary fuel for star formation and broadly trace the evolution of the cosmic SFR density from high redshifts to present day \citep[e.g.,][for a review]{Tacconi2020}.
   Dusty star-forming~\text{galaxies} (DSFGs) are a prominent manifestation of this intense, gas-rich, and dust-obscured phase of galaxy evolution, and have long been suspected to serve as the missing evolutionary bridge between the star-forming and quiescent phases of massive galaxy evolution \citep[e.g.,][]{Hodge2020, LeBail2024}. Since most star formation in DSFGs is dust-obscured, especially at cosmic noon, infrared (IR) emission from dust is often the only available tracer of SFR \citep[e.g.,][]{Whitaker2017, Shivaei2020, Solimano2026}. 
   Sub-kpc observations with the \textit{James Webb} Space Telescope (JWST) and the Atacama Large Millimeter/submillimeter Array (ALMA), which jointly trace the stellar, dust-obscured star-forming, and molecular gas components, are therefore essential for understanding the physical processes driving star formation in DSFGs \citep[e.g.,][]{Dessauges2025, Herrera2025, Herrera2026}. 
   
   Studies mapping galaxies at kpc scales have shown that star formation is tightly linked to the molecular gas content \citep[e.g.,][]{Bigiel2008}, giving rise to the Kennicutt--Schmidt relation \citep[KS;][]{Schmidt1959, Schmidt1963, Kennicutt1998}. This relation quantifies how efficiently molecular gas is converted into stars through a power law between the surface densities of SFR ($\Sigma_\mathrm{SFR}$) and cold gas ($\Sigma_\mathrm{gas}$), and it appears remarkably universal when measured from galaxy-integrated to kpc scales across a wide range of galaxy types and redshifts \citep[e.g.,][]{Leroy2008, Leroy2013, Freundlich2013}. It is typically expressed as $\Sigma_\mathrm{SFR}$\,$\propto$\,$\Sigma_\mathrm{gas}^k$, where $1$\,$\leq$\,$k$\,$\leq$\,$2$, depending on the tracers used and spatial scales considered \citep[e.g.,][]{Kennicutt2021}. 
   Observational tests of the KS relation rely on tracers of both the molecular gas reservoir and recent star formation. Because H$_2$ is extremely difficult to observe directly at the temperatures found in giant molecular clouds (GMCs), carbon monoxide (CO), the most abundant molecule after H$_2$, is widely used to trace molecular gas through its millimeter-wavelength rotational lines \citep[e.g.,][]{Carilli2013, Bolatto2013}. 
   H$\alpha$ is a widely used tracer of recent star formation, sensitive to newly formed stars with ages of $\sim$\,$0$--$10$~Myr \citep[e.g.,][]{Leroy2013}. However, rest-frame UV continuum and optical line emission from young stellar populations, including H$\alpha$, can be heavily dust-obscured. This absorbed radiation is reprocessed by dust and re-emitted in the total infrared (TIR; 3--1100\,$\upmu$m), making dust emission an important tracer of obscured star formation that has been shown to correlate closely with H$\alpha$ and UV emission \citep[e.g.,][]{Calzetti2013}. Dust, however, can be heated both by young stars from recent episodes of star formation and by the diffuse interstellar radiation field produced by the underlying quiescent stellar population \citep[e.g.,][]{Draine2007}. This makes TIR sensitive to star formation over extended timescales and can bias its use as an instantaneous SFR tracer. Nevertheless, in DSFGs where H$\alpha$ and UV emission are strongly attenuated, TIR remains a useful tracer of recent star formation \citep[][]{Kennicutt2012}.
   
   The KS relation has been extensively explored in the local universe, including at sub-galactic scales \citep[e.g.,][]{Leroy2013, Pessa2021, Sun2023}. Spatially resolved ($\sim$\,$100$~pc) observations of CO-to-H$\alpha$ flux ratios have revealed a universal decorrelation between molecular gas and young stars on GMC scales \citep[e.g.,][]{Schruba2010, Onodera2010}, enabling the underlying evolutionary timeline to be quantified and showing that GMC lifetimes are short, typically 10--30~Myr \citep[e.g.,][]{Chevance2020, Ramambason2026}. 
   Testing whether the same scale-dependent breakdown of the KS relation occurs in gas-rich, turbulent, and dust-obscured galaxies at cosmic noon remains observationally challenging, since it requires spatially resolved maps of both molecular gas and obscured star formation at sub-kpc scales. Beyond massive starburst galaxies, the cold molecular gas and dust properties of main-sequence (MS) galaxies ($<L^*$) can only be probed with very deep observations \citep[e.g.,][]{Molina2019, Arriagada2025} or by targeting strongly lensed systems \citep[e.g.,][]{Dessauges2019, Solimano2021, Catan2024}. 

   Strong gravitational lensing has proven to be a powerful tool to overcome these limitations, resolving the clumpy ISM of submillimeter galaxies at cosmic noon and beyond \citep[e.g.,][]{ALMA2015, Fujimoto2025}. When combined with high-resolution imaging and spectroscopy, such as those provided by JWST, strong lensing provides access to the internal physical scales that regulate galaxy evolution, from star-cluster scales \citep[parsec scales, e.g.,][]{Rigby2025, Messa2025, Adamo2025, Bradac2025, Claeyssens2026, Khullar2026} down to individual star candidates \citep[e.g.,][]{Welch2022, Furtak2024, Muller2025, Diego2026}. 
   ALMA studies leveraging strong gravitational lensing have enabled sub-kpc probes of the cold ISM at $z$\,$>$\,2, providing new insights into the structure, kinematics, and excitation conditions of star-forming regions \citep[e.g.,][]{Canameras2017, Rybak2020}. These observations suggest that the ISM is characterized by more compact dusty cores, elevated turbulence, and higher gas densities compared to local star-forming galaxies \citep[e.g.,][]{Messias2014, Rizzo2020, Vallini2021}. However, direct sub-kpc detections of the parent molecular clouds giving rise to UV-bright stellar clumps remain limited to a small number of strongly lensed systems, primarily at $z$\,$\sim$\,$1$, where successful CO detections in UV-selected sources, such as the Cosmic Snake and A521 galaxies, have reached resolutions down to $\sim$\,$30$~pc \citep[e.g.,][]{Dessauges2019, Dessauges2023, Nagy2023}. Yet, at $z$\,$>$\,2, such detailed spatially resolved studies remain scarce and largely limited to kpc scales \citep[e.g.,][]{Bethermin2023} or extreme starbursts \citep[$M_\ast$\,$>$\,$10^{10}\,{\rm M}_\odot$; SFR\,$\sim$\,$500$--$1000$\,${\rm M}_\odot\,{\rm yr}^{-1}$; e.g.,][]{Sharda2018, Vallini2024, Zanella2024, Vizgan2026}. 
   To build a complete picture of the star formation cycle during cosmic noon, it is therefore crucial to extend sub-kpc studies and tests of star formation laws across a broader range of ISM conditions and redshifts. 

   Among the remarkable strongly lensed DSFGs observed with ALMA, SMM\:J065837.6$-$555705, hereafter SMM\,J0658, is a uniquely extended disk-like galaxy at $z_{\mathrm{CO}}$\,=\,$2.7768$\,$\pm$\,0.0002 (this work), lensed by the Bullet Cluster \citep[$z$\,=\,0.296;][]{Tucker1998, Barrena2002} into three images. The high $\sim$\,$5$--$30\times$ lensing magnification provides a rare opportunity to map cold molecular gas and dust on sub-kpc scales at $z$\,$>$\,2. 
   We present a detailed analysis of the cold clumpy ISM based on high-quality ALMA follow-up observations, reaching angular and spectral resolutions down to 0\farcs2 and 3.2~km\,s$^{-1}$, respectively. We analyze two \text{$^{12}$C$^{16}$O(3--2)} ($\nu_{\rm rest}$\,=\,345.8~GHz) line cubes, hereafter \text{CO(3--2)}, at 0\farcs2 and 0\farcs6 resolutions, together with rest-frame 803\,$\upmu$m dust-continuum emission to test the KS relation over $\sim$200~pc--3~kpc scales. We also report the first submillimeter detection with ALMA of the faintest lensed image of SMM\,J0658, as well as evidence from ALMA and JWST observations for a lensed galaxy-pair candidate at $z$\,$\sim$\,$2.78$.

   The paper is structured as follows: in Section \ref{sec:obs}, we present the galaxy SMM\,J0658, the ALMA and JWST observations and data reduction. In Section~\ref{sec:methods}, we detail the methods of clump identification, source-plane reconstruction with the {\small LENSTOOL} code using a parametric strong-lensing model of the Bullet Cluster, and the derivation of physical parameters from the ALMA data. We present our results in Section \ref{sec:results} and compare them to previous studies in Section \ref{sec:discussion}. We summarize our main results and conclude in Section \ref{sec:conclusions}. 
   Throughout this study, we adopt a flat $\Lambda$CDM cosmology ($h$\,=\,0.7, $\Omega_\textrm{m}$\,=\,0.3, $\Omega_{\Lambda}$\,=\,0.7) and a Chabrier \citep[][]{Chabrier2003} initial mass function (IMF). At the Bullet Cluster redshift ($z$\,=\,$0.296$), $1\arcsec$ corresponds to 4.4~kpc, and at the source redshift ($z_{\mathrm{CO}}$\,=\,$2.7768$), $1\arcsec$ corresponds to 7.9~kpc.

\section{Observations and data reduction}
   \label{sec:obs}

\begin{figure*}[ht]
    \centering
	\includegraphics[trim=37 32 37 32, clip, width=0.99\textwidth]{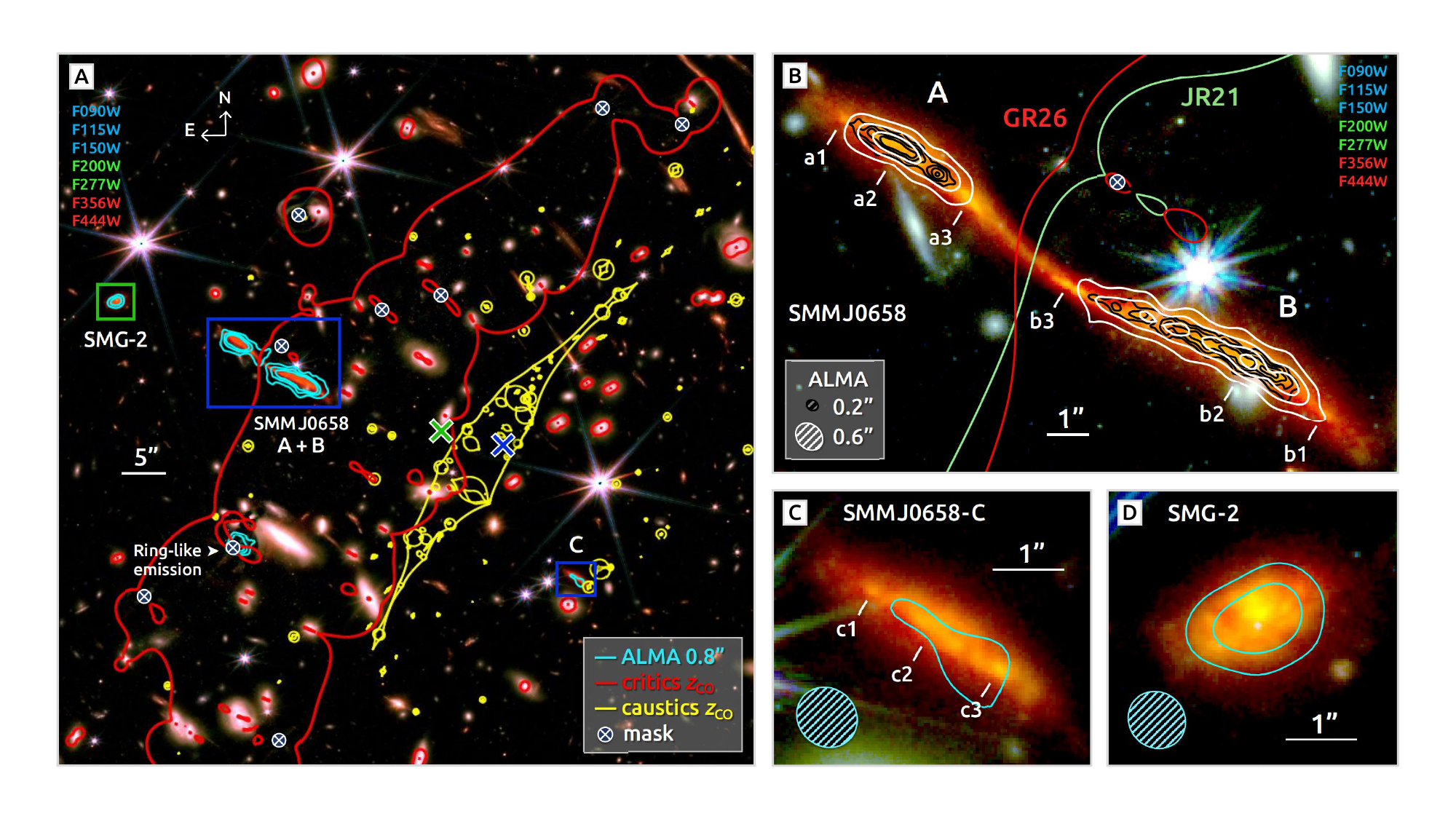}
    \caption{ 
    {\bf A:}~JWST/NIRCam short-wavelength F090W, F115W, F150W, and F200W imaging, combined with long-wavelength F277W, F356W, and F444W imaging of the Bullet Cluster, and overlaid with ALMA Band-3 contours from multi-frequency synthesis imaging at 0\farcs8 resolution (cyan contours at $3\sigma$, $5\sigma$, and $7\sigma$ significance levels). The critical lines predicted by the \citetalias{Rihtarsic2026} lens model are shown in red, and the caustics in yellow at the redshift of the \text{CO(3--2)} emission in SMM\,J0658, $z_{\rm CO}$\,=\,$2.7768$\,$\pm$\,$0.0002$. The two crosses mark the source-plane positions of SMG-2 (green cross) and SMM\,J0658 (blue cross), ray-traced at $z_{\rm CO}$. Their physical separation in the source plane is 60\,$\pm$\,5~kpc. The crossed circles indicate nine bright elliptical galaxies whose light has been masked. 
   {\bf B:}~Zoom-in on the two brightest lensed images (A and B) of SMM\,J0658. Contours of the \text{CO(3--2)} velocity-integrated flux are shown at 0\farcs2 resolution (black contours at $3\sigma$, $4\sigma$, $5\sigma$, $6\sigma$, and $7\sigma$ significance levels) and 0\farcs6 resolution (white contours at $3\sigma$, $6\sigma$, $9\sigma$, and $12\sigma$ levels). In each lensed image, we identify at least three similar morphological features in the NIRCam images. We compare the critical lines predicted by the JWST-based \citepalias{Rihtarsic2026} and pre-JWST \citepalias{Richard2021} lens models.  
   {\bf C:}~Zoom-in on the third lensed image of SMM\,J0658 (Image C). 
   {\bf D:}~Zoom-in on a submillimeter source at $z$\,$\sim$\,$2.78$ (SMG-2). 
   }
    \label{fig:jwst}
\end{figure*}

\subsection{Galaxy SMM\,J0658}
   SMM\,J0658 is a triply-imaged galaxy at $z_{\rm CO}$\,=\,$2.7768$\,$\pm$\,$0.0002$ (see Fig.~\ref{fig:jwst}), lensed by the massive Bullet Cluster (1E\,0657$-$56) at $z$\,=\,0.296 \citep{Tucker1998, Barrena2002}.
   The three images of SMM\,J0658 \citep[A, B, and C; following the notation of][]{Gonzalez2009} were first spectroscopically confirmed by \textit{Spitzer}/Infrared Spectrograph (IRS) observations (PID 496, PI~Gonzalez). \citet{Gonzalez2010} characterized SMM\,J0658 as a low-mass ($M_*$\,$\sim$\,$4\times10^{9}\ M_{\odot}$) and dusty starburst galaxy (SFR\,$\sim$\,100--150~${\rm M}_\odot\,\rm{yr}^{-1}$) with a far-IR luminosity $L_\mathrm{IR}$\,$\sim$\,$5\times 10^{11}\ {\rm L}_\odot$, comparable to that of L$_*$ galaxies at $z$\,$>$\,2 \citep[e.g.,][]{Forster2020}. The detection of polycyclic aromatic hydrocarbons (PAHs) and the absence of a clear \text{X-ray} counterpart suggest that the far-IR emission in SMM\,J0658 is primarily driven by star formation, although a contribution from an active galactic nucleus cannot be ruled out \citep[see][]{Gonzalez2010}. 
   
   The Bullet Cluster was observed by the HST/Advanced Camera for Surveys (ACS) in the F606W, F775W, and F850LP filters (PIDs 10200 and 10863, PIs Jones and Gonzalez), HST/Wide Field Camera~3 (WFC3) in the F814W filter (PID 11099, PI~Brada\v{c}), and more recently by the JWST/Near Infrared Camera (NIRCam) during the Cycle 3 General Observer (PID 4598, PIs~Brada\v{c}, Rihtar\v{s}i\v{c}, and Sawicki). In the rest-frame UV, SMM\,J0658 exhibits a very faint arc extending between the images~A and B in HST and NIRCam short-wavelength F090W, F115W, F150W, and F200W imaging. 
   VLT/MUSE 460--935\,nm observations ($2$\,h on-source, PID 094.A-0115, PI~Richard) do not show Ly$\alpha$ emission, consistent with significant dust obscuration. 
   
   \citet{Johansson2012} reported the first detections of \text{CO(1--0)} and \text{CO(3--2)} emission in SMM\,J0658 from observations with the Australia Telescope Compact Array (ATCA) at $\sim$15\arcsec\ angular resolution. We note that these ATCA observations did not resolve the CO double-horn profile, and that the measured redshift, $z$\,$=$\,$2.7793$\,$\pm$\,$0.0003$, may likely be biased toward the stronger peak on the receding side of the galaxy, which lies close to a critical line (see Fig.~\ref{fig:jwst}). \citet{Johansson2012} estimated the galaxy molecular gas content ($M_{\rm mol}$\,$\sim$\,$10^{10}\ {\rm M}_\odot$), and also complemented existing Herschel observations \citep[][]{Egami2010, Rex2010} with Atacama Pathfinder Experiment (APEX) SABOCA imaging of the rest-frame 350\,$\upmu$m continuum emission to quantify the galaxy dust properties ($T_{\rm dust}$\,$\sim$\,$33$\,$\pm$\,$5~\rm{K}$, and $M_{\rm dust}$\,$\sim$$10^{7}\,{\rm M}_\odot$). \citet[][hereafter, \citetalias{Motta2018}]{Motta2018} presented the first ALMA observations of SMM\,J0658 at 0\farcs6 resolution, identifying it as a rotating-disk system with a clumpy cold-gas morphology. 
   Here, we present a high 0\farcs2-resolution ALMA follow-up study of SMM\,J0658, using a new JWST-based strong-lensing model of the Bullet Cluster. 
   The lensing magnification and configuration make SMM\,J0658 an ideal system for spatially resolving cold molecular gas and dust properties at sub-kpc scales at $z$\,$\sim$\,2.78.

\subsection{ALMA observations and data reduction}
\label{sec:alma_dr}

\begin{figure*}[ht]
    \centering
    \includegraphics[width=0.99\textwidth]{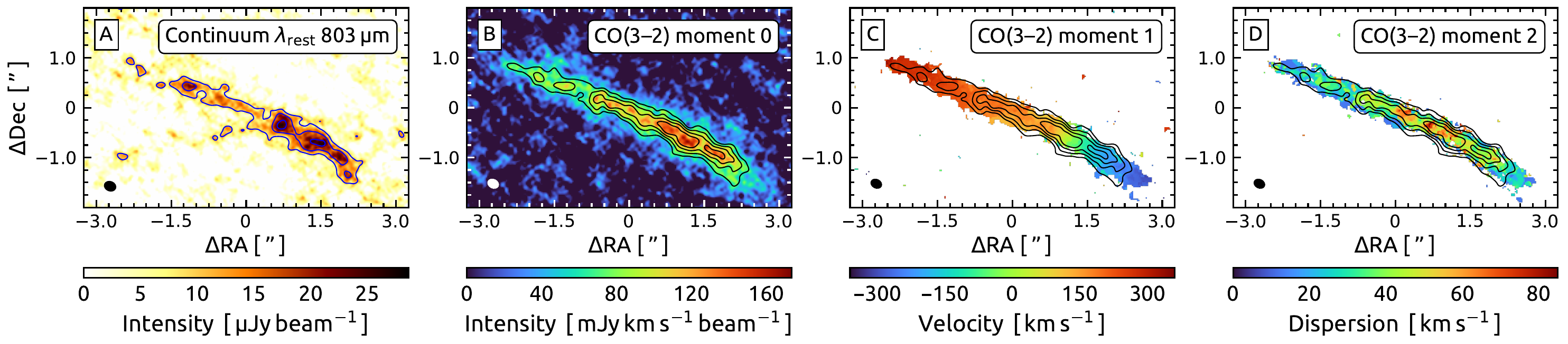}
    \caption{
    ALMA data at 0\farcs2 and $25.6\,{\rm km}\,{\rm s}^{-1}$ spectral resolution showing the highest magnified image of SMM\,J0658 (Image~B). In each map, the ellipse is the synthesized beam, the FOV is 6\farcs5\,$\times$\,4\arcsec, north is up, and east is left. 
    {\bf A:}~Rest-frame 803\,$\upmu$m continuum map. Contours at $2\sigma$, $3.5\sigma$, and $5\sigma$ significance are shown. 
    {\bf B:}~\text{CO(3--2)} velocity-integrated fluxes. \text{CO(3--2)} contours at $3\sigma$, $4\sigma$, $5\sigma$, and $6\sigma$ significance are shown in each moment map. 
    {\bf C:}~\text{CO(3–2)} rotation velocity map with a $3\sigma$ threshold, revealing a smooth disk-like rotation velocity profile.
    {\bf D:}~\text{CO(3–2)} velocity dispersion map with a $3\sigma$ threshold. Velocity dispersions in the mid-plane range between $\sim$\,50--80\,${\rm km}\,{\rm s}^{-1}$. 
    }
    \label{fig:B3HR_IP}
\end{figure*}

\begin{table*}[ht]
   \caption{Parameters of the ALMA Band-3 data used in the analysis, at angular resolutions of 0\farcs6 (2015.1.01559.S) and 0\farcs2 (2018.1.01754.S).}
   \label{tab:obs}
   \centering
   \begin{tabular}{p{4.3cm} p{3.8cm} p{3.8cm}}
     \hline\hline
     \noalign{\vskip 2pt}
     Program ID & 2015.1.01559.S & 2018.1.01754.S \\
     \noalign{\vskip 1pt}
     \hline
     \noalign{\vskip 2.5pt}
     SMM\,J0658 images covered & A and B & A, B, and C \\
     Observing frequency & 91.531 GHz & 91.531 GHz \\
     On-source time & 1.4 h & 6.9 h \\
     Array configuration(s) & C40-5 & C43-7; C43-8 \\
     Line-cube beam size\,\tablefootmark{a} & 0\farcs69 $\times$ 0\farcs53; PA = 42.8\degr & 0\farcs24 $\times$ 0\farcs19; PA = 63.4\degr \\
     Continuum beam size & 0\farcs69 $\times$ 0\farcs58; PA = 47.2\degr & 0\farcs25 $\times$ 0\farcs21; PA = 68.9\degr \\
     Maximum recoverable scale & $13\arcsec$ & 2\farcs8 \\
     Line RMS\,\tablefootmark{b} @25.6 km\,s$^{-1}$ & 180 $\upmu$Jy\,beam$^{-1}$~km\,s$^{-1}$ & 120 $\upmu$Jy\,beam$^{-1}$~km\,s$^{-1}$ \\
     Continuum RMS & 9.5 $\upmu$Jy\,beam$^{-1}$ & 4.4 $\upmu$Jy\,beam$^{-1}$ \\
     \noalign{\vskip 2pt}
     \hline
   \end{tabular}
   \tablefoot{ \label{notes1} 
         \tablefoottext{a}{Synthesized beam full-width at half maximum (FWHM) and position angle (PA), measured counter-clockwise from the north.} \\
         \tablefoottext{b}{The noise was measured over a spectral element of $25.6\,{\rm km}\,{\rm s}^{-1}$ in the CO(3--2) line cubes at 0\farcs2 and 0\farcs6 angular resolutions.}
         }
\end{table*}

   We used ALMA Band 3 to observe the CO\,($\nu$\,=\,$0$,~$J$\,=\,$3$\,$\rightarrow$\,$2$) rotational transition, redshifted to $91.558$~GHz for SMM\,J0658 at $z_\mathrm{CO}$\,=\,$2.7768$\,$\pm$\,$0.0002$. We also targeted the continuum emission centered at 98.8~GHz (rest-frame 803\,$\upmu$m) over a 7.5~GHz effective bandwidth. We used ALMA data from two observing campaigns: 0\farcs6 ALMA observations covering Images A and B, and a 0\farcs2 follow-up campaign with two pointings covering also Image C and including 0\farcs8 observations (PIDs 2015.1.01559.S and 2018.1.01754.S, PI~Motta). Table~\ref{tab:obs} lists the parameters of the ALMA data used in the present analysis. 

   The follow-up campaign was conducted in two pointings over five observing runs during Cycles 6 and 8. These observations used 45, 46, 20,\footnote{One of the observational execution blocks was obtained during the ALMA Return-to-Operations phase (after the COVID-19 shutdown), during which reduced antenna counts were permitted. We also include these data because they passed the QA2 quality assessment.} 40, and 47 antennas of the \text{12-m Array} in the C43-7 and C43-8 configurations, with maximum baselines of 3.7~km and 9.0~km, respectively. The total on-source time achieved was 6.9\,h.
   Due to the raw \textit{uv}-data volume ($\sim$0.8\,TB), the pipeline was run on the ALMA Observatory computing resources with the Common Astronomy Software Application \citep[][]{CASA2022} to restore the calibrated \textit{uv}-products (CASA version 6.2.1-7, pipeline version 2021.2.0.128). 
   We reduced the data using a total of four spectral windows, each with a bandwidth of 1.875~GHz and 1920 channels, corresponding to a 0.98\,MHz channel width (i.e., a velocity width of 3.2~km\,s$^{-1}$ at 91.531~GHz). The rest-frame 803\,$\upmu$m continuum emission with a total aggregate bandwidth of 7.5~GHz was imaged excluding channels presenting line emission. The \text{CO(3--2)} line cube was obtained after subtracting the continuum emission from the visibilities using the {\small{UVCONTSUB}} task. Imaging is done using the deconvolver routine of the {\small{TCLEAN}} task \citep[][]{Hogbom1974, Schwarz1978}. We restricted the imaging region to $800\times800$ pixels (20\arcsec$\times$\,20\arcsec) to speed up imaging, which is sufficient to encompass the lensed images A and B of SMM\,J0658. 
   The synthesized beam size is 0\farcs22 in the \text{CO(3--2)} line cube, obtained using Briggs weighting with a robust parameter of $R$\,=\,1.0, and 0\farcs23 in the continuum map (hereafter, 0\farcs2 data), obtained using natural weighting to enhance the signal-to-noise ratio (S/N) from the continuum emission, which is faint compared to CO. The final \text{CO(3--2)} line cube has a root-mean-square (RMS) noise of 260\,$\upmu$Jy\,beam$^{-1}$ per 3.2~km\,s$^{-1}$ channel in the original spectral resolution, and 120\,$\upmu$Jy\,beam$^{-1}$ per 25.6~km\,s$^{-1}$ after spectral averaging by a factor of 8. 
   In the rest-frame 803\,$\upmu$m continuum map, the noise level is 4.4\,$\upmu$Jy\,beam$^{-1}$. Figure~\ref{fig:B3HR_IP} presents the moment maps for Image~B produced using the {\small{IMMOMENTS}} task, excluding pixels below $3\sigma$ (360\,$\upmu$Jy\,beam$^{-1}$~km\,s$^{-1}$) for the \text{CO(3--2)} moment-1 and moment-2 maps.

   ALMA data at 0\farcs6 angular resolution were taken in two observing runs during Cycle 3, using 37 and 38 antennas of the \text{12-m Array} in the C40-5 configuration with a maximum baseline of 1.5~km. The total on-source time achieved in Band 3 was 1.4\,h. Calibrated \textit{uv} products were restored using the scripts provided to the PI (CASA version 5.4.0-68, pipeline version CASA54-P1-B r42030M following the QA2 report). 
   We re-reduced the data using a total of four spectral windows, each with a bandwidth of 1.875~GHz and 240 channels, corresponding to a 7.8\,MHz channel width (i.e., 25.6~km\,s$^{-1}$ velocity width at 91.531~GHz). The rest-frame 803\,$\upmu$m continuum emission with a total aggregate bandwidth of 7.5~GHz was imaged excluding channels presenting line emission. 
   The synthesized beam size is 0\farcs61 in the \text{CO(3--2)} line cube, obtained using Briggs weighting with robust parameter $R$\,=\,$1.0$, and 0\farcs63 in the continuum map, obtained using Briggs weighting with $R$\,=\,$0.5$ (hereafter, 0\farcs6 data). Different weightings were applied to maximize the S/N in each map, while matching the angular resolutions for the analysis. The final \text{CO(3--2)} line cube has a RMS noise of 180\,$\upmu$Jy\,beam$^{-1}$ per 25.6~km\,s$^{-1}$ channel width. In the rest-frame 803\,$\upmu$m continuum map, the noise level is 9.5\,$\upmu$Jy\,beam$^{-1}$. Moment maps of the \text{CO(3--2)} cube were produced using the CASA task {\small{IMMOMENTS}}, excluding pixels below $5\sigma$ (0.9\,mJy\,beam$^{-1}$~km\,s$^{-1}$) for the \text{moment-1} and \text{moment-2} maps (see Appendix~\ref{app:B3_IP}). 

\begin{figure}[ht]
	\includegraphics[width=0.99\columnwidth]{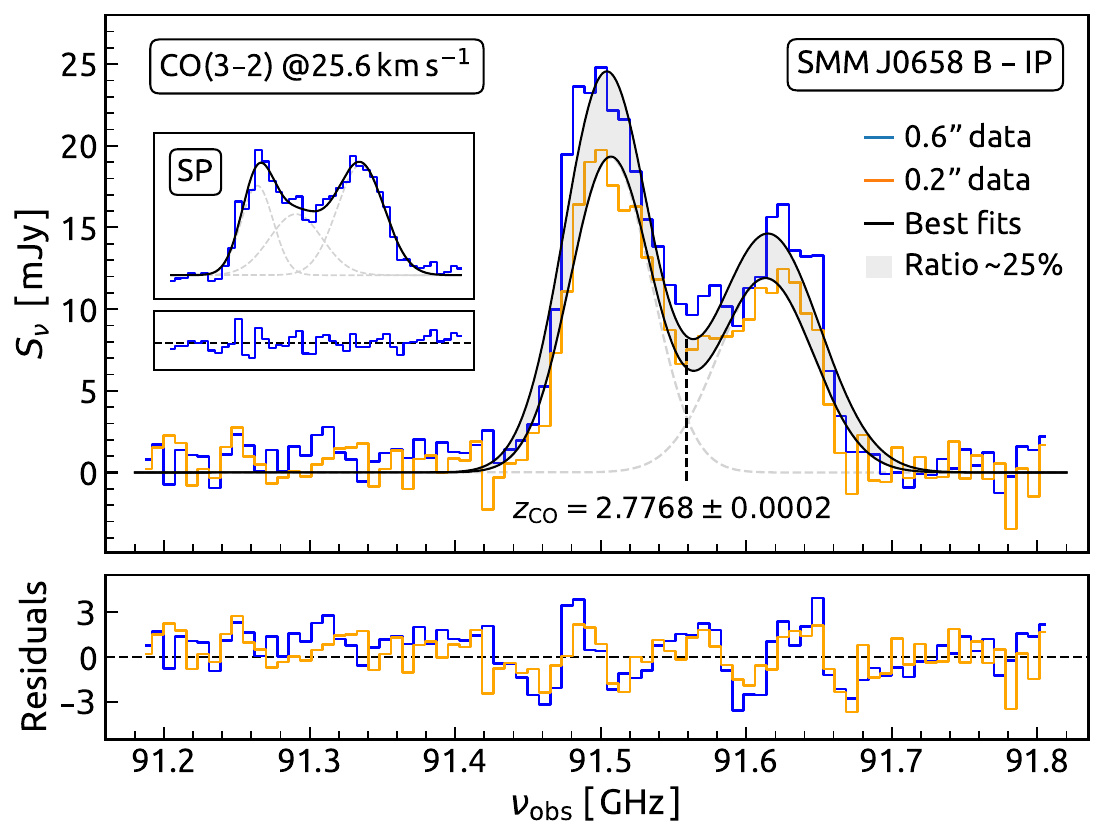}
    \caption{    
    The \text{CO(3--2)} line is detected at $91.558$~GHz in Image B of SMM\,J0658 in the continuum-subtracted line cubes at $0\farcs6$ and $0\farcs2$ angular resolution, with a common spectral resolution of $25.6\,{\rm km\,s^{-1}}$. The best-fit redshift is $z_{\rm CO}$\,=\,$2.7768\pm0.0002$. The gray shaded region corresponds to a missing diffuse-flux fraction of $25$\,$\pm$\,$4\%$ 
    at $0\farcs2$ compared to the $0\farcs6$ data.
    The inset shows the \text{CO(3--2)} spectrum in the $0\farcs6$ source-plane reconstructed line cube. The observed CO peak ratio of $1.7$\,$\pm$\,$0.3$ is attributed to the magnification gradient across Image~B.
    }
    \label{fig:CO32_spec}
\end{figure}

   Figure~\ref{fig:CO32_spec} shows the \text{CO(3--2)} spectra extracted from the brightest lensed image of SMM\,J0658 (Image~B, Fig.~\ref{fig:jwst}), where the S/N is highest, reaching an $8\sigma$ ($12\sigma$) detection at 0\farcs2 (0\farcs6) resolution. 
   The 0\farcs2 and 0\farcs6 \text{CO(3--2)} spectra, both measured at a spectral resolution of $25.6\,{\rm km}\,{\rm s}^{-1}$, are 25\,$\pm$\,4\% lower at 0\farcs2 compared to the 0\farcs6 data (see Fig.~\ref{fig:CO32_spec}). Since the largest angular scale in the source ($\sim$\,6\farcs5 for Image~B) is larger than the maximum recoverable scale limit in the 0\farcs2 observations ($\theta_{\rm MRS}$\,$\approx$\,2\farcs8; see Table~\ref{tab:obs}), this flux difference is consistent with missing the most extended emission in the 0\farcs2 data, but mainly affects galaxy-integrated measurements. 
   The line centroid, defined as the center of the \text{CO(3--2)} double-horn profile, yields $z_{\mathrm{CO}}$\,=\,2.7768\,$\pm$\,0.0002. This redshift is consistent between the spectra of Images~A and B at 0\farcs2 and 0\farcs6, obtained from independent data reductions and using either two- or three-Gaussian fits.
   We note a redshift difference of $\Delta z$\,=\,0.0011 relative to the \text{CO(3--2)} line redshift reported by \citetalias{Motta2018} from the same 0\farcs6 ALMA data, corresponding to a velocity offset of $\Delta v$\,=\,87\,${\rm km\,s^{-1}}$. This offset may arise because \citetalias{Motta2018} determined the redshift from the brightest CO peak rather than from the center of the double-horn profile. In any case, the offset is smaller than the apparent widths of 220--250\,${\rm km\,s^{-1}}$ FWHM measured for the individual CO line peaks.

\subsection{JWST/NIRCam observations}
   We used JWST/NIRCam short-wavelength (20~mas resolution) and long-wavelength (40~mas) imaging of the Bullet Cluster, observed in January 2025 as part of the GO program 4598. The Bullet Cluster was observed in eight filters (F090W, F115W, F150W, F200W, F277W, F356W, F410M, and F444W) with an exposure time of $\sim$6.4~ks in each filter. An RGB combination of NIRCam images is shown in Fig.~\ref{fig:jwst}. The observations used the {\small FULLBOX} six-point dither pattern to cover the gap between the two NIRCam modules. The images were processed following the The CAnadian NIRISS Unbiased Cluster Survey (CANUCS) data reduction pipeline described in \citet{Sarrouh2026}. This includes the modified stage 1 and stage 2 reductions with the STScI JWST pipeline, followed by redrizzling onto a 40~mas pixel grid using {\small GRIZLI} \citep{Brammer2023}. We refer the reader to \citet[][hereafter, \citetalias{Rihtarsic2026}]{Rihtarsic2026} for further details.

\section{Methods}
\label{sec:methods}

\subsection{Clump identification}
\label{sec:clump_id}
   At high angular resolution (typically 0\farcs1--0\farcs3 in the context of this study) achieved using extended ALMA 12-m Array configurations, interferometry filters out large-scale structures such as diffuse galaxy emission and emphasizes smaller, compact structures. This can subsequently lead to different interpretations of the resolved ISM \citep[e.g.,][]{Ivison2020, Faure2021}. In this work, we analyze the 0\farcs6 and 0\farcs2 ALMA data in parallel to test the KS relation and probe the spatial distribution of depletion times at kpc (0\farcs6) and sub-kpc (0\farcs2) scales. These datasets share the same observing frequency ($\nu_{\rm obs}$\,=\,$91.531\,{\rm GHz}$ in Band~3) and spectral resolution of $25.6\,{\rm km}\,{\rm s}^{-1}$ after spectral averaging by a factor of 8 in the 0\farcs2 data (see Table~\ref{tab:obs}).

\begin{figure*}[ht]
    \centering
	\includegraphics[width=0.95\textwidth]{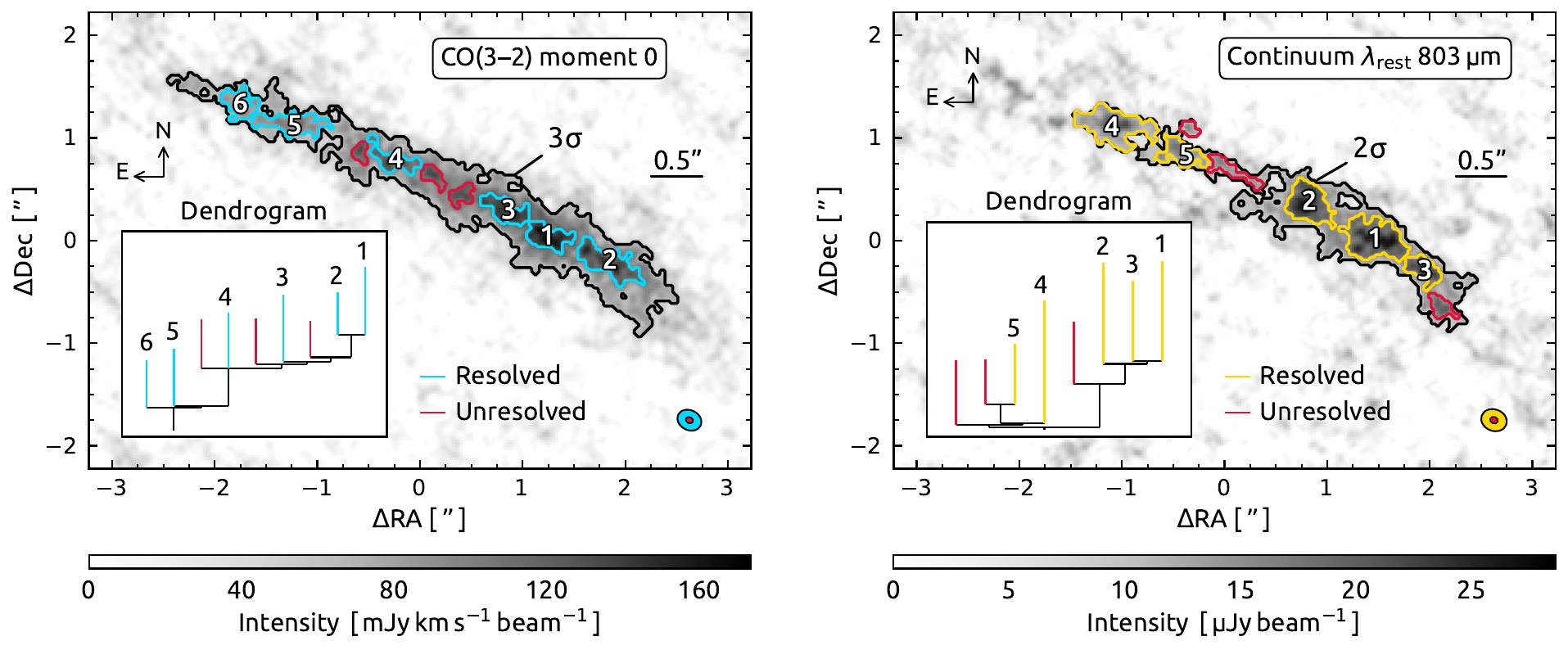}
    \caption{
    \textbf{Left:}~Giant molecular cloud complexes identified with {\small ASTRODENDRO} in the most magnified image of SMM\,J0658 (Image~B; Fig.~\ref{fig:jwst}) from the CO(3--2) velocity-integrated flux map at 0\farcs2 resolution. The synthesized beams used for the clump extraction are shown as ellipses (0\farcs2 in cyan for the resolved clumps; 0\farcs02 in red for the unresolved ones). The inset displays the corresponding dendrogram ordered by clump flux. 
    \textbf{Right:}~Obscured star-forming complexes identified in the rest-frame 803\,$\upmu$m continuum map at 0\farcs2, with the corresponding ellipses used for extraction, and dendrogram structure shown in the inset. 
    The spatial locations of the six resolved CO clumps and five dust-continuum clumps are compared in Fig.~\ref{fig:mu_map} (Panel~B).
    The clump properties are listed in Table~\ref{tab:GMC_prop}, and their source-plane locations are shown in Appendix~\ref{app:SP_dendro}.
    }
    \label{fig:dendro}
\end{figure*}

   To characterize the CO-traced GMC complexes and the cold dust-continuum clumps in the ALMA data, we produced dendrograms \citep[e.g.,][]{Rosolowsky2008, Goodman2009} using the publicly available {\small ASTRODENDRO}\footnote{\url{https://dendrograms.readthedocs.io/en/stable/}} code \citep[][]{Robitaille2019}. We constructed trees of flux isosurfaces, representing the hierarchy of emitting structures, referred to as leaves. We ran {\small ASTRODENDRO} on the \text{CO(3--2)} velocity-integrated flux maps and the rest-frame 803\,$\upmu$m continuum maps, identifying spatially resolved CO and dust clumps only in the $0\farcs2$ data. 
   Figure~\ref{fig:dendro} presents the resulting dendrograms, constructed by imposing (i)~a minimum detection threshold of $3\sigma$, above which fluxes were measured; (ii)~a minimum peak flux of $5\sigma$ for the leaves, defined as clumpy structures with no further substructure embedded within them; (iii)~a minimum flux contrast of $1\sigma$ between embedded structures; and (iv)~a minimum leaf area equal to the synthesized beam, i.e., $\mathcal{A}_{\rm leaf}$\,$\geq$\,$\frac{\pi}{4\ln 2}\,\theta_{\rm maj}\,\theta_{\rm min}$. 
   We tested the robustness of the extracted clumps by varying the initial thresholds and reducing the minimum leaf area by a factor of ten (i.e., 0\farcs02). At both 0\farcs2 and 0\farcs6 resolutions, {\small ASTRODENDRO} detected a few unresolved clumps (Fig.~\ref{fig:dendro}), but no additional substructure embedded within the spatially resolved clumps (see Appendix~\ref{app:0p6_dendro}). 
   Since unresolved clumps, i.e., those smaller than the synthesized beam according to criterion (iv), are difficult to interpret, we chose to exclude them from the analysis.
   We performed a 2D clump extraction for both ISM phases (CO gas and dust continuum) to ensure a consistent approach for the KS analysis. Taking the physical scales into account ($\sim$\,200~pc to 1~kpc; see Sect.~\ref{sec:sp_rec}), any CO clump identified in the 0\farcs2 \text{moment-0} map likely corresponds to a blend of multiple GMCs.

\subsection{Source-plane reconstruction}
\label{sec:src_rec}

\subsubsection{Lens model of the Bullet Cluster}
\label{sec:lens_model}
   The Bullet Cluster is a massive galaxy cluster merger at $z$\,=\,0.296 \citep[e.g.,][]{Tucker1998, Barrena2002, Clowe2006, Bradac2006}.
   We used an updated parametric strong-lensing model of the Bullet Cluster, presented in \citetalias{Rihtarsic2026}, made with the publicly available code {\small LENSTOOL}\footnote{\url{https://projets.lam.fr/projects/lenstool/wiki}} \citep{Kneib1996, Jullo2007, Jullo2010}. The model follows the previous parametric approach of \citet{Paraficz2016} and \citet[][hereafter, \citetalias{Richard2021}]{Richard2021} and includes new JWST/NIRCam imaging and NIRSpec spectroscopy (GO program 4598, PIs~Brada\v{c}, Rihtar\v{s}i\v{c}, Sawicki). 
   It is constrained by an expanded catalog of 135 secure multiple images from 27 distinct background galaxies, all with spectroscopic redshifts, representing a more than four-fold increase in spectroscopic constraints relative to previous studies \citep[e.g., \citetalias{Richard2021};][]{Cha2025}. The \citetalias{Rihtarsic2026} complete catalog of multiple images includes 199 multiple images from 73 lensed systems, belonging to 43 distinct galaxies. The addition of NIRSpec spectroscopy provides a wider and more uniform spatial coverage of spectroscopic systems compared to 
   previously, and extended the redshift range of multiple images from $2.8$\,$<$\,$z$\,$<$\,$3.5$ to $0.9$\,$<$\,$z$\,$<$\,$6.7$.
   The mass distribution is modeled with five large-scale and 219 galaxy-scale haloes, 213 of which follow luminosity-based scaling relations, and six galaxies are modeled separately.
   A fixed gas component derived from Chandra X-ray observations is also included \citep[][]{Bradac2006,Clowe2006,Randall2008}. All mass components adopt the Pseudo Isothermal Elliptical Mass Distributions \citep[PIEMDs;][]{Limousin2005, Eliasdottir2007}, except for the cluster gas. For a detailed description of the lens model, we refer the reader to \citetalias{Rihtarsic2026}.
   The substantial increase in spectroscopic constraints from JWST improves both the statistical and systematic uncertainties of the mass reconstruction, enabling more precise source-plane reconstruction.
   
   We derived from this lens model the predicted magnification ($\mu$) map at the source redshift. The mean absolute magnification measured over $3\sigma$ in the 0\farcs2 ALMA data is $\langle \mu_{\rm A} \rangle$\,=\,$9.0^{\smash{+0.8}}_{\smash{-0.1}}$ for Image~A and $\langle \mu_{\rm B} \rangle$\,=\,$19.3^{\smash{+2.2}}_{\smash{-1.5}}$ for Image~B (see Appendix~\ref{app:mu_map}). For Image~C, the mean absolute magnification measured in the NIRCam data is $\langle \mu_{\rm C} \rangle$\,=\,$5.5^{\smash{+0.2}}_{\smash{-0.3}}$. 
   We estimated the magnification by interpolating the convergence and shear maps from the best-fit lens model, and the uncertainties from 100 Bayesian posterior realizations. The $\mu$ values were measured directly from the best-fit magnification map, and their uncertainties were derived from the 16th and 84th percentiles of the posterior distribution.

\subsubsection{Reconstruction of SMM\,J0658}
\label{sec:sp_rec}
   We used custom {\small PYLENSTOOL} routines following the source-plane reconstruction methods implemented in {\small LENSTOOL} \citep[e.g., see][]{Sharma2018, Sharma2021, Patricio2019} to recover the intrinsic flux properties and morphology of SMM\,J0658 at $z_{\rm CO}$\,=\,2.7768. These routines ray-trace the image-plane data into source-plane maps from the best-fit parametric strong-lensing model.
   In practice, high magnification rates, steep magnification gradient, complex morphologies, and the presence of (partial) multiple images with different magnifications can make source-plane reconstruction challenging \citep[e.g.,][]{Patricio2019}.    
   Forward modeling methods addressing this exist \citep[e.g.,][]{Sharma2018, Sharma2021} but are beyond the scope of this study. 
   Throughout this work, we focus our analysis on Image~B, the most highly magnified image, which provides the best spatial resolution for studying the ISM of SMM\,J0658 down to scales of $\sim$200~pc. 
   
   We reconstructed the source-plane ALMA maps for Image~B at $z_{\rm CO}$\,=\,2.7768 using the data at $0\farcs6$ (pixel scale $0\farcs1$\,pix$^{-1}$) and $0\farcs2$ (pixel scale $0\farcs025$\,pix$^{-1}$). For both datasets, we applied an image-plane oversampling factor of 20 that corresponds approximately to the average magnification of Image~B, and adopted half of this value for the source-plane oversampling following \citet{Sharma2018}. These factors ensure adequate source-plane resolution, with each source-plane pixel sampled by at least one image-plane pixel. 
   The source-plane \text{CO(3--2)} flux and rest-frame 803\,$\upmu$m continuum maps were obtained by correcting for lensing distortions while preserving surface brightness, thereby recovering the intrinsic fluxes of SMM\,J0658. 
   In practice, the intrinsic fluxes can be derived either in the image plane (where the synthesized beam is constant) by applying a local magnification correction, or in the source plane (where the physical scale is constant) by ray-tracing the ALMA maps. Both approaches yield consistent results for clump-integrated measurements (see Appendix~\ref{app:mu_map}), because the magnification gradient across each clump is not significant. Furthermore, ratios of integrated quantities measured over the same regions can be derived in the image plane, since the magnification factor cancels out. When mapping these physical quantities spatially, however, it is more appropriate to work in the source plane for two reasons: (i)~the uniform physical scale allows surface-density maps to be derived consistently across the source; and (ii)~the intrinsic spatial locations and morphologies of the CO and dust clumps are recovered, enabling a better characterization of the source substructure. 
         
   In the source plane, the reconstructed beam varies across the source (see Fig.~\ref{fig:SP_beams}) and beams with smaller FWHM along directions of higher magnification probe smaller spatial scales. In the high-resolution 0\farcs2 ALMA data, we predict physical scales down to $\sim$200~pc at $z_{\rm CO}$ using the \citetalias{Rihtarsic2026} strong-lensing model, and an average physical resolution of $\sim$450~pc at $z_{\rm CO}$ that corresponds to a $\sim$15\% improvement compared to \citetalias{Richard2021}. Table~\ref{tab:sp_beams} lists the physical scale estimates for the 0\farcs2 and 0\farcs6 data. We also compared the source-plane reconstructed ALMA maps from the lensed images A and B of SMM\,J0658. Since the galaxy is seen entirely in these images, we expect the reconstructed source planes to exhibit the same morphology, with different effective resolutions due to the different magnifications. The model of \citetalias{Rihtarsic2026} produces source-plane morphologies of SMM\,J0658 that are more consistent between A and B than those derived using \citetalias{Richard2021}. We note here that this is expected due to the higher number of lensed galaxies with a spectroscopic redshift in \citetalias{Rihtarsic2026}, increased by a factor of four compared to previous models.

\begin{table}[ht]
   \caption{
   Physical scales reached in the source planes at $z_{\rm CO}$\,=\,2.7768 by ray-tracing Image~B of SMM\,J0658, using the \citetalias{Rihtarsic2026} lens model. 
   }
   \label{tab:sp_beams}
   \centering
   \small{
   \renewcommand{\arraystretch}{1.1}
   \begin{tabular}{rcc}
     \hline\hline
     \noalign{\vskip 2pt}
     & 0\farcs6 ALMA data & 0\farcs2 ALMA data \\
     \noalign{\vskip 0.5pt}
     \hline
     \noalign{\vskip 2pt}
     $\langle \theta_{\rm cube} \rangle$ [pc\,$\times$~pc]  & 1730\,$\times$\,770  & 590\,$\times$\,280 \\
     $\langle \theta_{\rm cont} \rangle$ [pc\,$\times$~pc]  & 1770\,$\times$\,790  & 620\,$\times$\,290 \\
     $\theta_{\rm \,mean}$  [pc]                            & 1270                 & 450 \\
     $\theta_{\rm \,b,min}$ [pc]                            & 520                  & 200 \\
     \noalign{\vskip 1pt}
     \hline
   \end{tabular}
   }
   \tablefoot{ Values are rounded to the nearest ten parsecs. $\langle \theta_{\rm cube} \rangle$ and $\langle \theta_{\rm cont} \rangle$ are the averaged source-plane synthesized beams for the \text{CO(3--2)} line cube and rest-frame 803\,$\upmu$m continuum map, respectively. $\theta_{\rm mean}$ is the mean physical resolution in the $\tau_{\rm dep}$ maps shown in Fig.~\ref{fig:KS_GR26}, computed by averaging the major and minor axes of both $\langle \theta_{\rm cube} \rangle$ and $\langle \theta_{\rm cont} \rangle$. $\theta_{\rm b,min}$ is the smallest source-plane angular scale along the arc (beam minor axis). }
\end{table}
   
   We also reconstructed the source-plane \text{CO(3--2)} line cubes for Image~B of SMM\,J0658. Each spectral channel map of $25.6\, {\rm km}\,{\rm s}^{-1}$ velocity width was ray-traced to the source plane at $z_{\rm CO}$\,=\,2.7768, and the resulting source-plane channel maps were stacked into a new cube. Figure~\ref{fig:CO32_spec} shows that the observed CO peak ratio of 1.7\,$\pm$\,0.3 in the double-horn profile is absent in the reconstructed source-plane spectrum, suggesting that the peak ratio arises from the magnification gradient across the arc (with the highest CO peak corresponding to the side of the galaxy closest to the critical line; see Fig.~\ref{fig:jwst}) rather than from an intrinsic inhomogeneity in the \text{CO(3--2)} emission in SMM\,J0658. The differential magnification may also explain the asymmetry in the image-plane \text{CO(3--2)} rotation velocity field (see Fig.~\ref{fig:B3HR_IP}; moment~1), by amplifying the receding side of the disk that lies closer to the critical line.

\subsection{Determination of physical quantities}
\label{sec:phys_quantities}

\subsubsection{Star formation rate surface density}

   We estimated the total stellar mass, SFR, and dust attenuation coefficient of SMM\,J0658 by modeling its rest-frame $\sim$0.5--800~$\upmu$m spectral energy distribution (SED) with the {\small CIGALE} code \citep{Boquien2019}, including the ALMA and NIRCam flux densities. A description of the SED model is provided in Appendix~\ref{app:SED}. 
   Assuming that the rest-frame 803~$\upmu$m dust-continuum emission traces the spatial distribution of star formation, the SED-inferred SFR was used to calibrate the 0\farcs2 and 0\farcs6 continuum maps, $\mathcal{C}_{803}$, and derive SFR surface density maps as 
   \begin{equation}
   \label{eq:sb_corr}
       \Sigma_{\rm SFR} = \alpha_{\rm SED}\ \mathcal{C}_{803}\ {\rm cos}\,i\, \left( D_{\rm A}^{2} \, \frac{\Omega_{\rm beam}}{\Omega_{\rm pixel}} \right)^{-1} \, {\rm M}_\odot\,{\rm yr}^{-1}\,{\rm kpc}^{-2} \,,
   \end{equation}
   where $\alpha_{\rm SED}$ is the scaling factor used to recover the observed total SFR within $3\sigma$ continuum emission, $i=70\pm5^\circ$ is the best-fit disk inclination measured for Image~B in the source-plane maps, $D_A^2$ is the physical area at the source redshift, and $\Omega_{\rm beam}/\Omega_{\rm pixel}$ is the number of pixels per synthesized beam, converting the map units from flux density per beam to flux density per pixel.

\subsubsection{Molecular gas mass surface density}

   We used the CO(3--2) line detected at $91.558$~GHz with ALMA as a tracer of the cold, H$_2$-dominated molecular gas. 
   The CO line luminosity, $L'_{\rm CO}$ [$\rm{K~km\,s^{-1}~pc^2}$], can be expressed for a source of any size in terms of the total line flux \citep{Solomon1997} as
   \begin{equation}
      \label{eq:L'_CO}
      L'_{\rm CO} = \frac{3.25 \times 10^7}{(1+z)^3} \, \left( \frac{S_{\rm CO}\Delta V}{\rm{Jy~km\,s^{-1}}} \right) \left( \frac{\nu_{\rm obs}}{\rm GHz} \right)^{-2} \left( \frac{D_L}{\rm Mpc} \right)^2 \, {\rm L}_\odot \,,
   \end{equation}
   where $z$ is the CO line redshift, $S_{\rm CO}\Delta V$ the velocity-integrated flux density, $\nu_{\rm obs}$ the observed CO line frequency, and $D_L$ the luminosity distance to the source. 
   The \text{CO(3--2)} line luminosity, $L'_{\rm CO(3-2)}$, was converted into molecular gas mass as
   \begin{equation}
      \label{eq:M_gas}
       M_{\rm mol} = \left( \, \frac{\alpha_{\rm CO}}{{(\rm K~km\,s^{-1}~pc^2})^{-1}} \right)
       \left( \frac{L'_{\rm CO(3-2)}/r_{31}}{\rm K~km\,s^{-1}~pc^2} \right) \ {\rm M}_\odot \,,
   \end{equation}
   where $r_{31}$\,=\,$L'_{\smash{\rm CO(3-2)}}/L'_{\smash{\rm CO(1-0)}}$ is the CO line luminosity ratio, and $\alpha_{\rm CO}$ is the CO-to-H$_2$ conversion factor. We used $r_{31}$\,=\,0.56, as measured by \citet{Johansson2012} from ATCA observations. This ratio is slightly below the average value of 0.66 reported for submillimeter galaxies in statistical surveys \citep[see][Table 2]{Carilli2013}. 
   For the CO-to-H$2$ conversion factor, we adopted an upper limit of $\alpha_{\rm CO}$\,=\,$3.0$~$[{\rm K~km\,s^{-1}\,pc^2}]^{-1}$, previously determined by \citetalias{Motta2018} from the dynamical mass constraint $M_{\smash{\rm mol}} \leq M_{\smash{\rm dyn}} - M_{\smash{\ast}}$. We used our revised SED-based stellar mass estimate of $(4.0\pm1.0)\times10^{10}\ {\rm M}_\odot$, disk inclination angle $i$, and a half-light radius of $R_{\smash{1/2}}$\,=\,$1.6$\,$\pm$\,$0.2$~kpc. We verified that the virial masses of the CO clumps, inferred from their mean velocity dispersions of $\sim$25--60\,$\pm$\,5~${\rm km}\,{\rm s}^{-1}$ measured in the 0\farcs2 CO moment-2 map, are consistent with gravitationally bound structures and with the adopted upper limit on $\alpha_{\rm CO}$. 
   Molecular gas surface density maps were then derived as: 
   \begin{equation}
       \Sigma_{\rm mol} = M_{\rm mol}\ {\rm cos}\,i\, \left( D_{\rm A}^{2} \, \frac{\Omega_{\rm beam}}{\Omega_{\rm pixel}} \right)^{-1} \, {\rm M}_\odot\,{\rm pc}^{-2} \,.
   \end{equation}

   We computed the galaxy-integrated $\Sigma_{\rm mol}$ and $\Sigma_{\rm SFR}$ from the 0\farcs6 ALMA data over the galaxy half-light area, $\mathcal{A}_{\smash{\rm 1/2}}$\,=\,$\pi R^{\smash{2}}_{\smash{1/2}}$, accounting for the disk inclination. We then derived spatially resolved maps of $\Sigma_{\rm mol}$ and $\Sigma_{\rm SFR}$ at 0\farcs2 and 0\farcs6 by calibrating the source-plane \text{CO(3--2)} moment-0 and continuum maps, respectively, so that their integrated values over an effective aperture of area $\mathcal{A}_{\smash{\rm 1/2}}$ recover the global estimates. This effective aperture follows the source-plane CO and continuum morphologies.

\section{Results}
\label{sec:results}

\subsection{Global properties of SMM\,J0658}
\label{res:gal_prop}

   We infer from the SED modeling (Fig.~\ref{fig:sed}) a galaxy-integrated stellar mass of $M_\ast$\,=\,$(4.0\pm1.0)\times10^{10}\ {\rm M}_\odot$, a dust attenuation of $A_V$\,=\,$3.3\pm0.8$~mag, consistent with the value previously derived by \citet{Gonzalez2010}, and a total SFR\,=\,$47\pm12\ {\rm M}_\odot\,{\rm yr}^{-1}$, as listed in Table~\ref{tab:J0658_prop}. The significant dust attenuation, together with the very faint optical counterparts, suggests that SMM\,J0658 is a heavily obscured DSFG \citep[e.g.,][]{Casey2014}. 
   
\begin{figure}[ht]
    \includegraphics[width=0.99\columnwidth]{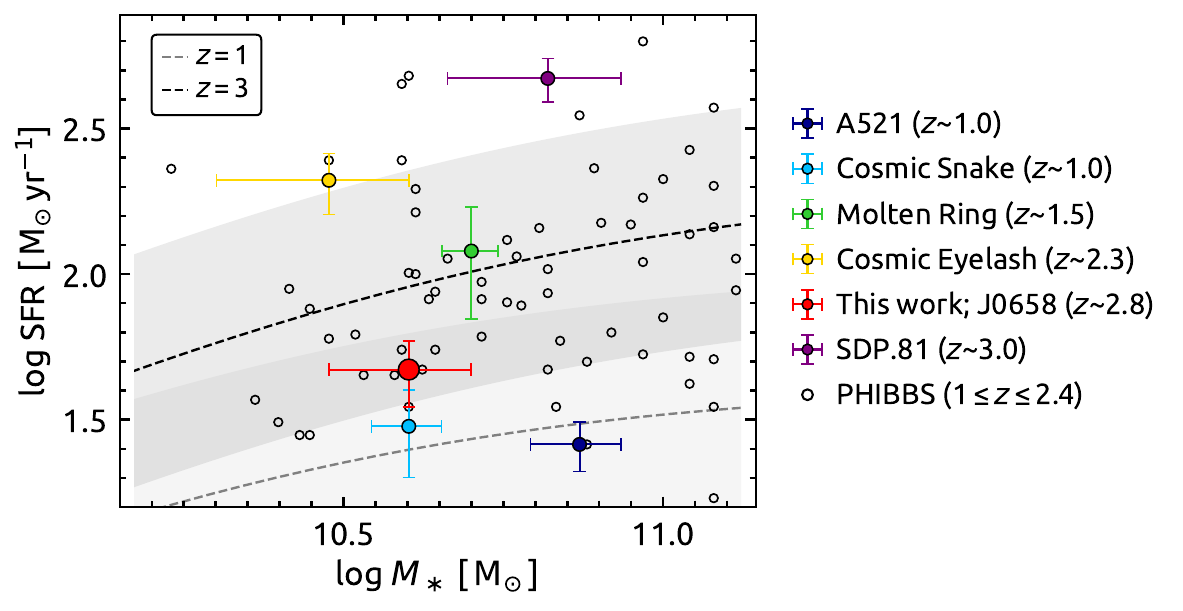}
    \caption{
    Location of SMM\,J0658 in the $M_\ast$--\,SFR plane compared to other lensed galaxies at 1\,$\leq$\,$z$\,$\leq$\,3 \citep[][]{Swinbank2010, Tacconi2013, Dye2015, Sharda2018, DiazSanchez2021, Dessauges2019, Nagy2023}. Main sequence models at $z$\,=\,1 and $z$\,=\,3 are shown \citep{Popesso2023}, with shaded regions corresponding to the $\pm\,0.4\,\mathrm{dex}$ boundary \citep{Genzel2014}.
    }
    \label{fig:MS_plane}
\end{figure}

   We derive an upper limit of $M_{\rm mol} \leq (5.0\pm1.3) \times 10^{10}\ {\rm M}_\odot$ from the data, corresponding to $\Sigma_{\rm mol} \leq (1.13\pm0.29)\times10^3\,{\rm M}_\odot\,{\rm pc}^{-2}$.  Using the revised stellar mass estimate from the SED modeling, we derive $\Sigma_{\rm SFR}=1.1\pm0.3\,{\rm M}_\odot\,{\rm yr}^{-1}\,{\rm kpc}^{-2}$, and an upper limit on the gas fraction of $f_{\rm gas} \leq M_{\rm gas}/(M_{\rm gas}+M_\ast)=60\pm20\%$, consistent with the high molecular gas fraction observed in galaxies at cosmic noon \citep[][]{Forster2020}.
   The corresponding upper limit on the gas depletion time is $\tau_{\rm dep} \leq 1.06\pm0.27\ {\rm Gyr}$. This relatively long depletion time is mainly driven by the lower SFR inferred from our SED modeling, which includes both NIRCam photometry and ALMA flux densities, compared to previous SED-based estimates \citep[e.g.,][\citetalias{Motta2018}]{Gonzalez2009, Johansson2012}. 
   SMM\,J0658 lies within the main-sequence population at $z_{\rm CO}$, as shown in Fig.~\ref{fig:MS_plane}, where we compare its location in the \text{$M_\ast$--SFR} plane with other lensed galaxies at cosmic noon with stellar masses between $10^{10}$\,$\lesssim$\,$M_\ast$\,$\lesssim$\,$ 10^{11}\,{\rm M}_\odot$.

\begin{table}[ht]
    \caption{Global properties of SMM\,J0658.}
    \label{tab:J0658_prop}
    \centering
    \renewcommand{\arraystretch}{1.15}
    \begin{tabular}{ll}
        \hline\hline
        \noalign{\vskip 1pt}
        Quantity & SMM\,J0658 \\
        \noalign{\vskip 0.75pt}
        \hline
        \noalign{\vskip 2pt}
        $z_{\rm CO}$ & $2.7768\pm0.0002$ \\
        $A_{\rm V}^{\textcolor{blue}{\dagger}}\ [{\rm mag}]$ & $3.3\pm0.8$ \\ 
        $M_\ast^{\textcolor{blue}{\dagger}}\ [10^{10}\,{\rm M}_\odot]$ & $4.0\pm1.0$ \\ 
        ${\rm SFR}^{\textcolor{blue}{\dagger}}\ [{\rm M}_\odot\,{\rm yr}^{-1}]$ & $47\pm12$ \\
        $M_{\rm mol}^{\textcolor{red}{\ast}}\ [10^{10}\,{\rm M}_\odot]$ & $5.0\pm1.3$ \\
        $f_{\rm gas}^{\textcolor{red}{\ast}}\ [\%]$ & $60\pm20$ \\
        $\Sigma_{\rm mol}^{\textcolor{red}{\ast}}\ [{\rm M}_\odot\,{\rm pc}^{-2}]$ & $1130\pm290$ \\
        $\Sigma_{\rm SFR}\ [{\rm M}_\odot\,{\rm yr}^{-1}\,{\rm kpc}^{-2}]$ & $1.1\pm0.3$ \\
        $\tau_{\rm dep}^{\textcolor{red}{\ast}}\ [{\rm Gyr}]$ & $1.06\pm0.27$ \\
        ${\rm SFE}\ [{\rm Gyr}^{-1}]$ & $0.94\pm0.24$ \\
        \noalign{\vskip 2pt}
        \hline
    \end{tabular}
    \tablefoot{ $\textcolor{blue}{\dagger}$~Values inferred from the galaxy SED modeling (see Fig.~\ref{fig:sed}). 
    $\textcolor{red}{\ast}$~Upper-limit estimates resulting from the adopted upper limit on $\alpha_{\rm CO}$. }
\end{table}

\subsection{Spatially resolved CO and dust clumps}
\label{sec:res_clumps}

   The ALMA data reveal a clear spatial decorrelation between the cold molecular gas and dust distributions in SMM\,J0658 (Fig.\,\ref{fig:B3HR_IP}). At 0\farcs6 angular resolution, the dust-continuum emission is more compact than the molecular gas, consistent with star formation occurring preferentially in the inner regions of the galaxy, while the molecular gas is clumpier and more extended across the disk. At the higher 0\farcs2 resolution, both the molecular gas and dust appear clumpy, but their clumps are not spatially coincident. Using {\small ASTRODENDRO}, we identify a total of 11 spatially resolved clumps in the 0\farcs2 ALMA maps, including six CO clumps and five dust clumps (Fig.~\ref{fig:dendro}). We measure the intrinsic clump properties in the 0\farcs2 source-plane maps (Table~\ref{tab:GMC_prop}), after reconstructing the morphology of the clumps at $z_{\rm CO}$ (Appendix~\ref{app:SP_dendro}). The intrinsic clump-emission areas used for the measurements correspond to effective radii ranging from $\sim$230 to 480~pc at $z_{\rm CO}$. We find massive ($M_{\rm mol}$\,$\sim$\,$10^9\pm0.3~{\rm M}_\odot$) star-forming clumps, characterized by $\Sigma_{\rm mol}$\,$\sim$\,1300--2300~${\rm M}_\odot\,{\rm pc}^{-2}$ and $\Sigma_{\rm SFR}$\,$\sim$\,0.5--2.3~${\rm M}_\odot\,{\rm yr}^{-1}\,{\rm kpc}^{-2}$. This corresponds to upper limits on the molecular gas depletion time of $\sim$0.82--2.56~Gyr, as listed in Table~\ref{tab:GMC_prop}.

\subsection{Sub-kpc test of the Kennicutt--Schmidt relation} 
\label{sec:res_morpho}

\begin{figure*}[ht]
	\includegraphics[width=0.99\textwidth]{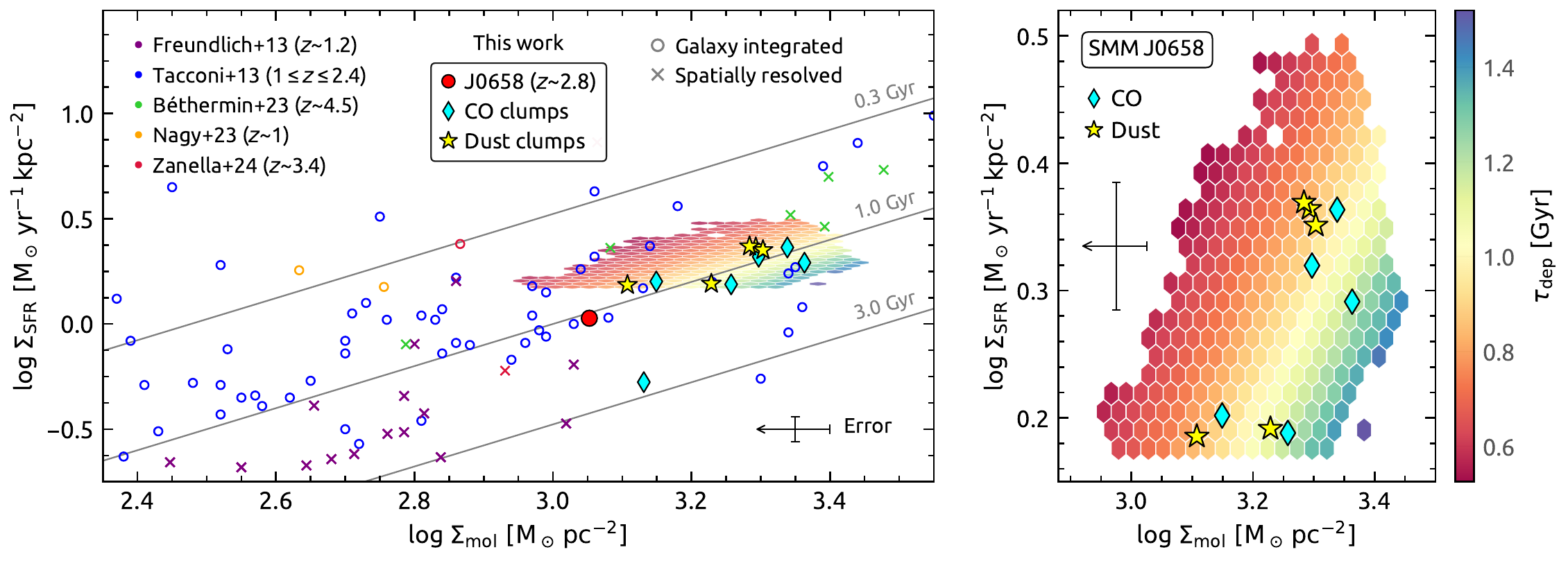}
    \caption{ 
    {\bf Left:}~Star-forming clumps of SMM\,J0658 in the Kennicutt--Schmidt plane, compared with other studies at kpc scales for $z$\,$\sim$\,1--4.5 \citep[][]{Genzel2010, Tacconi2013, Freundlich2013, Bethermin2023}, and sub-kpc scales for $z$\,$\sim$\,1 and $z$\,$\sim$\,3.4 \citep[][respectively]{Nagy2023, Zanella2024}. Galaxy-integrated measurements are shown as circles, while spatially resolved measurements appear as crosses. 
    The global value for SMM\,J0658 is indicated by a red circle, and the clump-integrated measurements by diamonds for the CO clumps and stars for the dust clumps. 
    The uncertainty is indicated by an error-bar cross in each panel for SMM\,J0658, the CO and dust clumps, with the arrow marking the mass upper limit. For clarity, error bars are shown only for our data, but are comparable to those of the other studies. 
    All $\Sigma_{\rm mol}$ measurements are based on CO line detections, except for \citet{Zanella2024}, which used the [C\,\textsc{ii}] 158\,$\upmu$m line. Gray solid lines indicate constant molecular gas depletion times, $\tau_{\rm dep}$\,$=$\,$\Sigma_{\rm mol}/\Sigma_{\rm SFR}$, of 0.3, 1.0, and 3.0~Gyr, while the colored background shows $\tau_{\rm dep}$ continuously across the plane for SMM\,J0658. 
    {\bf Right:}~Zoom-in on the pixel-to-pixel $\Sigma_{\rm SFR}$--$\Sigma_{\rm mol}$ diagnostics for SMM\,J0658, shown as a hexbin density map and corresponding to the colored background region shown in the left panel. Colors indicate $\tau_{\rm dep}$, computed after applying a $3\sigma$ threshold on the $\Sigma_{\rm mol}$ and $\Sigma_{\rm SFR}$ maps to reduce~scatter. 
    }
    \label{fig:KS_plane}
\end{figure*}

   We studied the $\Sigma_{\rm mol}$--$\Sigma_{\rm SFR}$ correlation separately in the 0\farcs2 ALMA data for CO- and dust-dominated clumps. 
   Figure~\ref{fig:KS_plane} reveals that the clump-integrated $\Sigma_{\rm mol}$ and $\Sigma_{\rm SFR}$ measurements do not lead to significant deviations from the KS relation, except for one CO clump with a very low dust content (i.e., higher $\tau_{\rm dep}$\,$\sim$2.6~Gyr). Without this outlier, we find $\Sigma_{\rm SFR}$\,$\propto$\,$\Sigma_{\smash{\rm mol}}^{\smash{k}}$ with $k$\,$\sim$1.0--1.2. The mean depletion time is $\sim$1.28~Gyr for the CO clumps and $\sim$0.90~Gyr for the dust clumps, consistent with higher star-formation efficiency (SFE\,=\,$\tau_{\smash{\rm dep}}^{\smash{-1}}$) in regions dominated by obscured star formation. 
   At the pixel-to-pixel level, the $\Sigma_{\rm mol}$--$\Sigma_{\rm SFR}$ plane shows depletion times spanning $\tau_{\rm dep}$\,$\sim$\,0.5--1.5~Gyr (see Fig.~\ref{fig:KS_plane}; right-hand panel). This motivated us to probe the spatial distribution of the molecular gas depletion time across the source-plane reconstructed galaxy. 
   
   \begin{table}[ht]
   \caption{
   Intrinsic physical properties of the spatially resolved clumps identified in the ALMA Band-3 data at 0\farcs2 angular resolution.
   }
   \label{tab:GMC_prop}
   \renewcommand{\arraystretch}{1.25}
   \centering
   \resizebox{0.99\columnwidth}{!}{%
   \begin{tabular}{rccccc}
     \hline \hline
     \noalign{\vskip 1pt}
     ID & $\langle \, \mu \, \rangle$ & $\mu^{-1}R_{\rm eff}$ & $\Sigma_{\rm mol}$ & $\Sigma_{\rm SFR}$ & $\tau_{\rm dep}$ \\
        & & {\small [pc]} & {\small [$10^{3}\, {\rm M}_\odot\, \rm{pc}^{-2}$]} & {\small [${\rm M}_\odot\, \rm{yr}^{-1}\, \rm{kpc}^{-2}$]} & {\small [Gyr]} \\
     \noalign{\vskip 2pt}
     \hline
     \noalign{\vskip 0.6pt}
     \multicolumn{6}{c}{ CO(3--2) } \\ 
     \noalign{\vskip 0.2pt}
     \hline 
     \noalign{\vskip 2pt}
         \cyancircle{1} & $16.6\,^{\smash{+2.5}}_{\smash{-1.8}}$ & 
         $311$ & $2.3\,\pm\,0.6$ & $2.0\,\pm\,0.5$ & $1.18\,\pm\,0.30$ \\
         \cyancircle{2} & $13.7\,^{\smash{+2.0}}_{\smash{-1.3}}$ & 
         $439$ & $2.0\,\pm\,0.5$ & $2.1\,\pm\,0.5$ & $0.95\,\pm\,0.24$ \\
         \cyancircle{3} & $19.6\,^{\smash{+3.4}}_{\smash{-2.3}}$ & 
         $286$ & $2.2\,\pm\,0.5$ & $2.3\,\pm\,0.6$ & $0.94\,\pm\,0.24$ \\
         \cyancircle{4} & $24.8\,^{\smash{+3.8}}_{\smash{-2.1}}$ & 
         $233$ & $1.8\,\pm\,0.5$ & $1.5\,\pm\,0.4$ & $1.17\,\pm\,0.29$ \\ 
         \cyancircle{5} & $21.4\,^{\smash{+3.1}}_{\smash{-1.3}}$ & 
         $316$ & $1.4\,\pm\,0.4$ & $1.6\,\pm\,0.4$ & $0.89\,\pm\,0.22$ \\
         \cyancircle{6} & $18.6\,^{\smash{+3.6}}_{\smash{-1.7}}$ & 
         $280$ & $1.4\,\pm\,0.3$ & $0.5\,\pm\,0.1$ & $2.56\,\pm\,0.64$ \\ 
     \noalign{\vskip 2pt}
     \hline
     \noalign{\vskip 0.6pt}
     \multicolumn{6}{c}{ Dust-continuum } \\ 
     \noalign{\vskip 0.2pt}
     \hline
     \noalign{\vskip 2pt}
         \goldcircle{1} & $15.8\,^{\smash{+2.4}}_{\smash{-1.7}}$ & 
         $482$ & $2.0\,\pm\,0.5$ & $2.3\,\pm\,0.6$ & $0.85\,\pm\,0.21$ \\ 
         \goldcircle{2} & $20.1\,^{\smash{+3.6}}_{\smash{-2.4}}$ & 
         $359$ & $1.9\,\pm\,0.5$ & $2.3\,\pm\,0.6$ & $0.82\,\pm\,0.21$ \\ 
         \goldcircle{3} & $13.2\,^{\smash{+1.8}}_{\smash{-1.3}}$ & 
         $314$ & $2.0\,\pm\,0.5$ & $2.2\,\pm\,0.6$ & $0.90\,\pm\,0.23$ \\ 
         \goldcircle{4} & $22.4\,^{\smash{+3.1}}_{\smash{-1.3}}$ & 
         $408$ & $1.3\,\pm\,0.3$ & $1.5\,\pm\,0.4$ & $0.84\,\pm\,0.21$ \\
         \goldcircle{5} & $24.7\,^{\smash{+3.2}}_{\smash{-1.6}}$ & 
         $250$ & $1.7\,\pm\,0.4$ & $1.6\,\pm\,0.4$ & $1.09\,\pm\,0.27$ \\
     \noalign{\vskip 2pt}
     \hline
     \end{tabular}
     }
     \tablefoot{ \label{tab2} 
          Clump-integrated properties measured in the source plane at $z_{\rm CO}$, accounting for the magnification gradient across the clumps through a pixel-by-pixel correction. Clump IDs follow the notation used in Fig.~\ref{fig:dendro}. The effective radius of the source-plane area used for the measurements at $z_{\mathrm{CO}}$ is defined as $R_{\rm eff}$\,=\,$({A/\pi})^{1/2}$. $R_{\rm eff}$ is fixed by the adopted measurement area and is therefore reported without uncertainty. 
          }
\end{table}

\begin{figure*}[ht]
    \centering
    \includegraphics[width=0.99\textwidth]{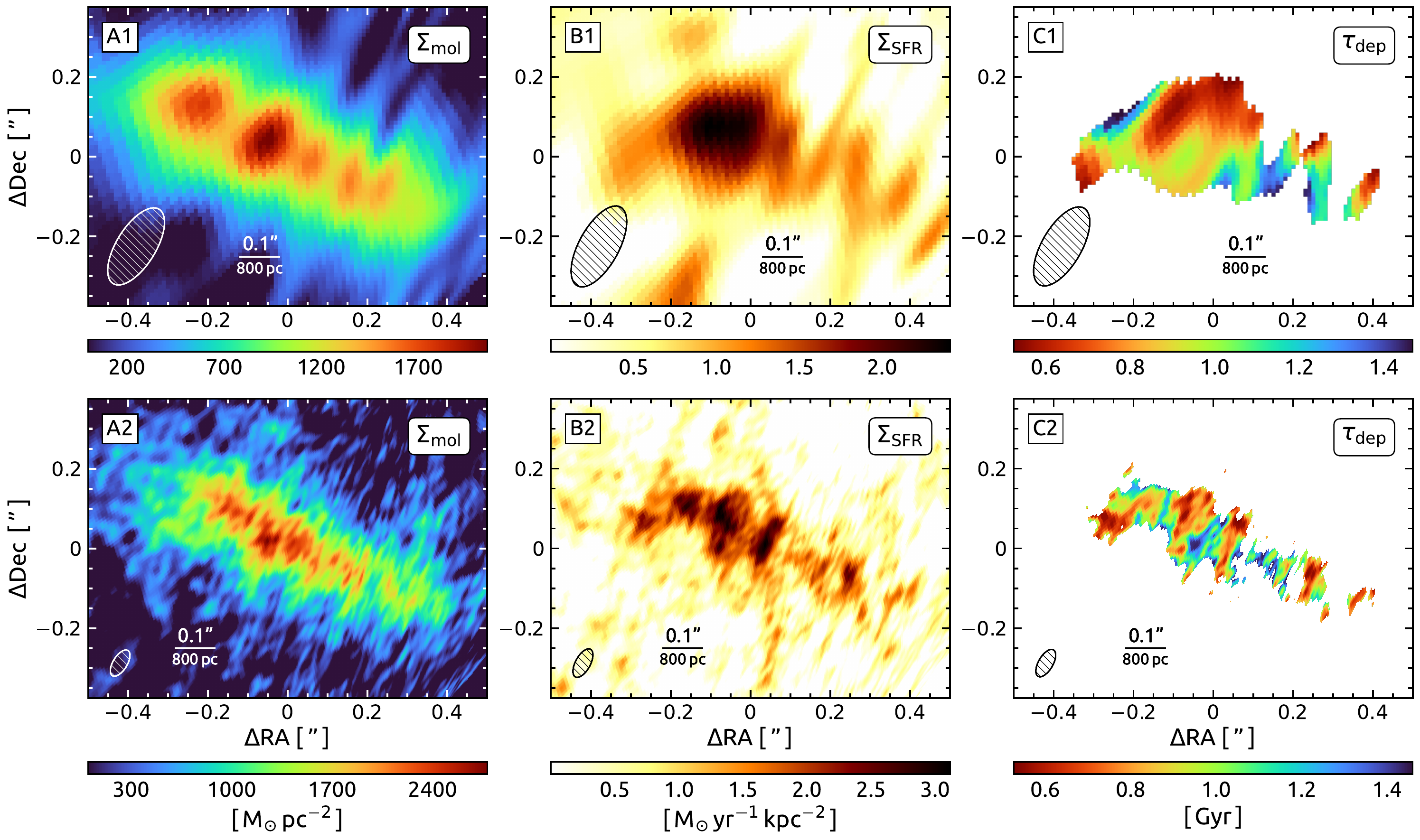}
    \caption{
    Source-plane reconstructed ALMA maps of Image~B of SMM\,J0658 at $z_{\rm CO}$\,=\,$2.7768$, using the \citetalias{Rihtarsic2026} parametric strong-lensing model. 
    {\bf A:}~Source-plane molecular gas mass surface density maps, $\Sigma_{\rm mol}$, derived from the CO(3--2) velocity-integrated flux maps at 0\farcs6 (\textbf{A1}) and 0\farcs2 (\textbf{A2}). 
    {\bf B:}~Source-plane SFR surface density maps, $\Sigma_{\rm SFR}$, derived from the rest-frame 803\,$\upmu$m continuum maps at 0\farcs6 (\textbf{B1}) and 0\farcs2 (\textbf{B2}). 
    {\bf C:}~Source-plane maps of molecular gas depletion time at $z_{\rm CO}$ with a 0.1~dex uncertainty, computed as $\tau_{\rm dep}$\,=\,$\Sigma_{\rm mol}/\Sigma_{\rm SFR}$ at 0\farcs6 (\textbf{C1}) and 0\farcs2 (\textbf{C2}), after applying a $3\sigma$ threshold to reduce scatter. 
    In each map, the ellipse is the averaged source-plane reconstructed synthesized beam. North is up and east is left. 
    }
    \label{fig:KS_GR26}
\end{figure*}

   Figure~\ref{fig:KS_GR26} presents source-plane maps of $\Sigma_{\rm mol}$, $\Sigma_{\rm SFR}$, and $\tau_{\rm dep}$ derived from the 0\farcs6 and 0\farcs2 ALMA data. A $3\sigma$ threshold was applied to the $\tau_{\rm dep}$ maps to reduce scatter. We note that, since the dust continuum is fainter than the CO emission, the morphology of the depletion time maps closely follows the dust-continuum distribution. 
   The highest $\Sigma_{\rm mol}$ and $\Sigma_{\rm SFR}$ regions remain broadly co-spatial at both angular resolutions. 
   The 0\farcs2 data resolve the broad central region with $\tau_{\rm dep}$\,$\sim$\,0.6~Gyr seen in the 0\farcs6 data into several compact sub-regions of star formation. 
   We compare these source-plane $\tau_{\rm dep}$ maps obtained from the JWST-based lens model of \citetalias{Rihtarsic2026} with those obtained using the \citetalias{Richard2021} model in Appendix~\ref{app:sp_tau}. We discuss the origin of the scale dependence of the molecular gas depletion time in Sect.~\ref{sec:disc_taudep}.

\subsection{Evidence for a lensed galaxy pair at $z$\,$\sim$\,$2.78$}
\label{sec:gal_pair}
   \citetalias{Motta2018} reported the discovery of a fourth image of SMM\,J0658 (hereafter, \text{SMG-2}) using the same 0\farcs6 ALMA data. However, the parametric strong-lensing models of the Bullet Cluster used in this study \citepalias{Richard2021, Rihtarsic2026} predict \text{SMG-2} to be a singly imaged source, whether constrained by the 0\farcs2 and 0\farcs6 ALMA data or by the NIRCam observations, suggesting that \text{SMG-2} and SMM\,J0658 are distinct~\text{galaxies}. We note that the lens model used in \citetalias{Motta2018} did not predict any critical line between \text{SMG-2} and the nearest lensed image of SMM\,J0658, which already favored the interpretation of two distinct sources. Using the JWST-based \citetalias{Rihtarsic2026} model, we find a projected physical separation between SMM\,J0658 and \text{SMG-2} of $60$\,$\pm$\,$5$~kpc in the source plane at $z_{\rm CO}$ (see Fig.~\ref{fig:jwst}), which is consistent with typical galaxy separations observed in interacting pairs and merger candidates \citep[e.g.,][]{Conselice2014, Kaviraj2025}. Moreover, ALMA Cycle 11 observations using the \text{12-m} Array in Band~4 (2024.1.00015.S, PI~Cornil-Ba\"{i}otto) detected the \text{CO(5--4)} line at 152.6~GHz in \text{SMG-2} at $z$\,=\,2.7753\,$\pm$\,0.0005, with 8$\sigma$ significance in the \text{CO(5--4)} moment-0 map, and also confirmed the revised CO redshift of SMM\,J0658. We do not present a detailed analysis of the \text{Band-4} data, as they are beyond the scope of this study. The redshift offset between SMM\,J0658 and \text{SMG-2} corresponds to rest-frame recessional velocities of $\Delta \nu_i = c \, \Delta z / (1+z_i)\approx$~120~km\,s$^{-1}$, which is on the order of galaxy peculiar velocities, as well as typical velocity offsets in galaxy pairs and merger candidates. In the NIRCam images shown in Fig.~\ref{fig:jwst}, three identical substructures are identified in each of the SMM\,J0658 images \citepalias{Rihtarsic2026}, while \text{SMG-2} appears different in morphology, surface brightness distribution, color, and symmetry, reinforcing the hypothesis of two distinct sources at $z$\,$\sim$\,$2.78$.

\subsection{Additional submillimeter detections}
   The ALMA data reveal two additional submillimeter sources in the field. 
   We report the first ALMA detection of the faintest lensed image of SMM\,J0658 (Image~C; see Fig.~\ref{fig:jwst}), covered by the second pointing from the follow-up campaign (Table~\ref{tab:obs}). However, Image~C is detected only at $3\sigma$ in the 0\farcs8 multi-frequency synthesis imaging (i.e., combining the CO and continuum emission), and is undetected in the 0\farcs2 data. 
   ALMA also resolves the morphology of continuum emission located near one of the brightest cluster galaxies in the Bullet Cluster (see Fig.\,\ref{fig:jwst}), previously reported by \citet{Johansson2012}, revealing a ring-like morphology. However, no spectral line is detected in this source. 
   Further spectroscopic follow-up is required to determine whether this emission is associated with the cluster galaxy or with a background/foreground source.

\section{Discussion}
\label{sec:discussion}

\subsection{Efficiency of star formation}
\label{sec:disc_litterature}

   At the current global SFR, the cold molecular gas reservoir of SMM\,J0658 would be exhausted within $\tau_{\rm dep}$\,$\leq$\,$1.06\pm0.27$~Gyr, assuming continuous star formation. This depletion time corresponds to a SFE of $0.94$\,$\pm$\,$0.24\ {\rm Gyr}^{-1}$, consistent with typical values for main-sequence galaxies at 1\,$<$\,$z$\,$<$\,3 \citep[e.g.,][]{Nagy2023, Arriagada2025}, and lower than those of extreme starburst systems, such as SDP.81 at $z$\,$\sim$\,3 \citep[e.g.,][]{Dye2015, Sharda2018}. 
   However, since galaxies evolve secularly, accrete and expel gas, and interact with one another over cosmic time, their SFRs and SFEs are expected to vary considerably, from episodes of intense star formation to quiescent phases \citep{Boquien2019}. 
   The physical conditions inferred for the $0\farcs2$ ALMA clumps in SMM\,J0658 are broadly consistent with those of main-sequence galaxies. Their $\Sigma_{\rm SFR}$ (see Table~\ref{tab:GMC_prop}) are lower than those measured in the SDP.81 clumps studied by \citet{Sharda2018}, and the pixel-to-pixel KS diagnostics (see Fig.~\ref{fig:KS_plane}) yield depletion times consistent with those of $\sim$0.2--1.2~Gyr measured in lensed $z$\,$\sim$\,1 main-sequence galaxies on sub-kpc scales \citep{Nagy2023}. 
   The shortest dust-clump depletion times also overlap with the upper end of the depletion time range measured in $z$\,$\sim$\,4.5 main-sequence galaxies from the ALPINE--CRISTAL survey, $\tau_{\rm dep}$\,$\sim$0.29--0.84~Gyr \citep{Bethermin2023, Accard2025}. 
   However, these last comparisons should be interpreted with caution, since resolved KS measurements at high redshift can be highly sensitive to the adopted gas-mass calibration. For example, in [C\,\textsc{ii}]-based studies, adopting a constant $\alpha_{\rm [C\,\textsc{ii}]}$ can yield depletion times of $\sim$0.5--1~Gyr of main-sequence galaxies, whereas a $\Sigma_{\rm [C\,\textsc{ii}]}$-dependent conversion can produce much steeper KS slopes and push the densest regions into the starburst regime, with $\tau_{\rm dep}$\,$<$\,0.1~Gyr \citep[e.g.,][]{Zanella2018, Zanella2024, Vallini2025}.

   The spatially resolved $\tau_{\rm dep}$ maps reveal localized regions with depletion times lower than $\sim$700~Myr (see Fig.~\ref{fig:KS_GR26}), comparable to the $\sim$0.2--1.2~Gyr depletion times typically measured in $z$\,$\sim$\,$1$ main-sequence galaxies \citep[e.g.,][]{Nagy2023}, but longer than the $\sim$100~Myr depletion times measured in nearby starburst galaxies at $\sim$\,200~pc resolution \citep[e.g.,][]{Fisher2022}. These regions of enhanced SFE may arise from local disk instabilities that drive rapid gas inflows \citep[e.g.,][]{Fisher2022}, an evolutionary stage in which GMC collapse triggers star formation before stellar feedback has dispersed the parent clouds \citep[e.g.,][]{Kruijssen2019, Chevance2020}, a possible interaction with \text{SMG-2} \citep[e.g.,][]{Boquien2011b, Dye2015, Tacconi2018}, or a combination of these mechanisms. The smooth velocity field measured by ALMA (see Fig.~\ref{fig:B3HR_IP}) disfavors strong global kinematic perturbations, but does not exclude local disk instabilities.
   We found a physical separation of 60\,$\pm$\,5~kpc in the source plane and a rest-frame recessional velocity offset of $\sim$\,120~km\,s$^{-1}$ of \text{SMG-2} relative to SMM\,J0658, both consistent with typical galaxy separations observed in interacting pairs and merger candidates \citep[e.g.,][]{Conselice2014, Kaviraj2025}. We further note that SMM\,J0658 and \text{SMG-2} appear morphologically quiescent in NIRCam imaging, with no clear tidal features, possibly indicating an early interaction phase. Simulations and observations of galaxy pairs involving massive galaxies ($M_\ast$\,$>$\,$10^{10}\,{\rm M}_\odot$) have shown that over $\sim$40\% of pairs with physical separations in the range 50--100~kpc and with $\lesssim$\,300~km\,s$^{-1}$ velocity differences are likely to merge on timescales of $\sim$2~Gyr \citep[][]{Kitzbichler2008, Patton2024}. The stellar mass ratio of SMM\,J0658 to \text{SMG-2} \citepalias[$M_\ast$\,=\,$3.2^{\smash{+3.8}}_{\smash{-1.8}}$\,$\times$\,$10^{10}$\,${\rm M}_\odot$ in a 4.2~kpc aperture;][]{Motta2018} is $\sim$1.3, placing the system within the major merger regime \citep[e.g.,][]{Kaviraj2025}. While interaction with \text{SMG-2} could enhance the efficiency of star formation in SMM\,J0658 \citep[e.g.,][]{DiMatteo2007, Moreno2021}, the strong variations in gas depletion time are resolved on sub-kpc scales, where the evolutionary cycling of molecular clouds and star-forming regions is expected to play a central role.

\subsection{Scale dependence of the molecular gas depletion time}
\label{sec:disc_taudep}
   The strong $\tau_{\rm dep}$ variations across SMM\,J0658 are driven by the spatial decorrelation of CO and dust at sub-kpc scales (see Fig.~\ref{fig:KS_GR26}). CO and H$\alpha$ observations of local star-forming galaxies down to $\lesssim$\,100~pc scales have shown that this spatial decorrelation is more likely associated with the time evolution of distinct, individual regions of star formation than with observational noise, stellar feedback, or the drift of young stars away from their parent clouds \citep{Schruba2010}. 
   In our case, observational noise is unlikely to dominate given the high S/N of the ALMA data, while stellar feedback is expected to affect much smaller scales ($\lesssim$\,10~pc) than those resolved here. Although dust-based SFR tracers are sensitive to older stellar populations ($\lesssim$\,100~Myr) than H$\alpha$ ($\lesssim$\,10~Myr), their luminosity-weighted stellar ages have comparable values of $\sim$5~Myr for TIR emission and $\sim$3~Myr for H$\alpha$ \citep[e.g.,][]{Kennicutt2012, Leroy2013}. We therefore do not expect stellar drift from parent clouds to dominate the observed gas--dust spatial decorrelation in the ALMA~data. 
   
   We investigated the scale dependence of the depletion time in SMM\,J0658 by measuring CO-to-dust flux ratios in the 0\farcs2 source-plane ALMA maps, where the physical scale is uniform. We first ray-traced the clump locations into the source plane (see Appendix~\ref{app:SP_dendro}). We then characterized the correlation between GMC complexes and obscured star-forming regions by placing apertures centered on the CO or dust clumps, and measuring how the enclosed CO-to-dust flux ratios are elevated or suppressed relative to the galaxy-scale average as the aperture size changes \citep[e.g.,][]{Schruba2010, Kruijssen2014}. 
    
\begin{figure}[ht]
    \centering
    \includegraphics[width=0.99\columnwidth]{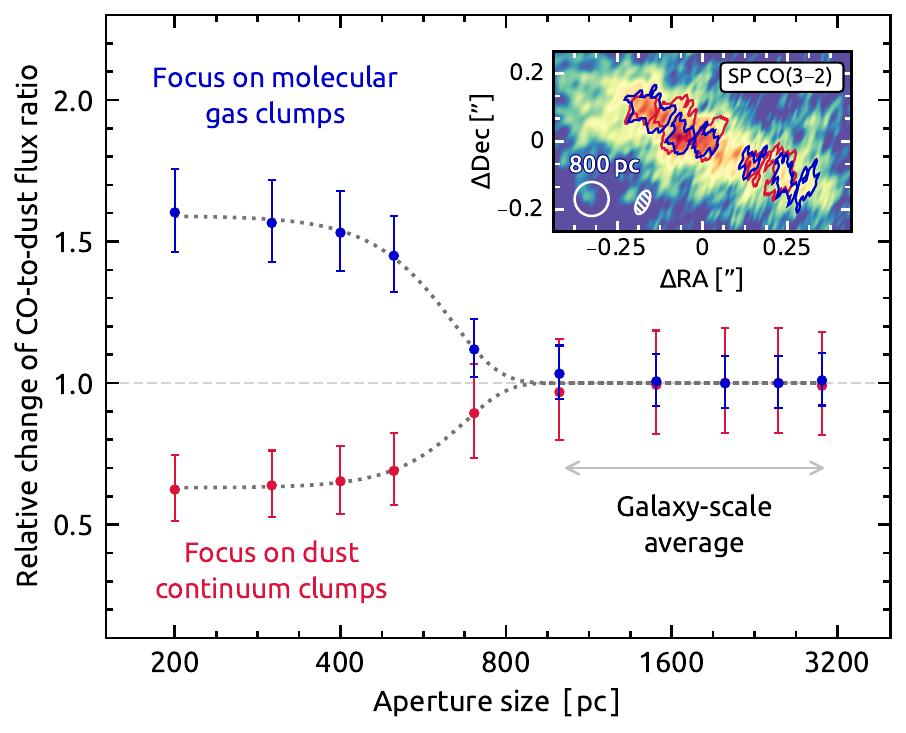}
    \caption{
     Spatial correlation between the \text{CO(3--2)} and rest-frame~803\,$\upmu$m continuum emission in SMM\,J0658 in the 0\farcs2 ALMA data, tracing the scale dependence of the molecular gas depletion time, $\tau_{\rm dep}$\,=\,$\Sigma_{\rm mol}/\Sigma_{\rm SFR}$. 
     CO-to-dust ratios are measured in source-plane apertures ranging from 200~pc to 3.2~kpc in diameter, centered on CO clumps tracing molecular gas (blue) or continuum clumps tracing obscured star formation (red). 
     An exponential convergence profile is fitted to each branch (dashed curves) to estimate the KS relation breakdown scale, where the ratios diverge from the galaxy-average value. 
     The inset shows the source-plane \text{CO(3--2)} integrated intensity map at 0\farcs2 resolution, with the resolved clumps, an 800~pc aperture, and the average source-plane beam.
    }
    \label{fig:KS_fork}
\end{figure}

   Figure~\ref{fig:KS_fork} presents the spatial correlation of molecular gas and dust-obscured star-forming regions from 200~pc to 3.2~kpc. We find a tuning-fork behavior similar to local studies using CO, H$\alpha$, and 21~$\upmu$m emission tracers, with a breakdown scale of 0.8\,$\pm$\,0.1~kpc in SMM\,J0658, consistent with typical values of $\sim$500~pc\,--\,1~kpc measured in local galaxies \citep[e.g.,][]{Chevance2020, Kim2022, Ramambason2026}. These local studies used observations down to $\sim$50~pc scales, while the mean physical resolution in our 0\farcs2 source-plane maps is $\sim$450~pc. At $z$\,$\sim$\,1, \citet{Nagy2023} report a larger breakdown scale of $\sim$6~kpc for a main-sequence star-forming galaxy, using ALMA observations down to 30~pc in the source plane and FUV emission instead of H$\alpha$. It remains unclear whether this difference is driven by the stellar tracer adopted, galaxy disk inclination, or ISM conditions. SMM\,J0658 is also a highly inclined galaxy, yet we recover the same tuning-fork behavior using rest-frame 803~$\upmu$m dust continuum as a star formation tracer. 
   
   For kpc-scale apertures, the CO-to-dust ratio converges toward the galaxy-scale average, while for apertures smaller than $\sim$0.8~kpc, the two separated branches diverge by up to $\sim$0.2~dex in depletion times (see Fig.~\ref{fig:KS_fork}). An asymptotic plateau is reached for apertures smaller than $\lesssim$\,300~pc, possibly due to the 0\farcs2 data resolution limit. 
   Using CO-to-dust flux ratios, we find a smaller depletion-time scatter than local studies based on CO-to-H$\alpha$ ratios. A similar reduction in scatter is observed in local galaxies when using 21~$\upmu$m emission, which traces dust-embedded feedback phases, instead of H$\alpha$ \citep[][]{Ramambason2026}. However, local studies typically focus on nearly face-on galaxies, and SMM\,J0658 is a highly inclined disk-like system. Projection effects may therefore blend distinct clumps along the line of sight and reduce the apparent separation between molecular gas- and dust-dominated clumps. 
   For apertures smaller than the breakdown scale, dust-dominated clumps are systematically offset toward shorter depletion times, or equivalently higher SFEs, consistent with active dust-obscured star-forming regions. In contrast, CO-dominated clumps are offset toward longer depletion times, consistent with large molecular gas reservoirs that may either not yet have entered efficient gravitational collapse or not yet be significantly dispersed by stellar feedback. Our results are consistent with a scale-dependent breakdown of the KS relation driven by the resolution of individual regions tracing distinct evolutionary stages of the star formation cycle in SMM\,J0658. 

\subsection{GMC lifetime estimates in SMM\,J0658}
\label{sec:GMC_lifetime} 

   \citet{Kruijssen2014} formalize the physical scales at which star-formation relations break down in galaxies, when the gas and radiation tracers associated with star formation no longer correlate. They show that the depletion-time scatter is sensitive to the physical sizes and timescales involved in the star-formation process, including its duration and the molecular cloud lifetime. Their methodology has been successfully applied to a wide range of local galaxies \citep[e.g.,][]{Kruijssen2018, Chevance2020}, yielding GMC lifetimes of 10--30~Myr across a range of galaxy environments \citep{Kim2022}, and has been extended to comparisons between different stellar tracers \citep{Ramambason2026}. It remains uncertain, however, whether high-$z$ GMCs have similar lifetimes. In the following, we discuss the possibility of applying this method to the ALMA data and its implications for GMC lifetimes in SMM\,J0658. 
   
   Following Eq.~5 of \citet{Kruijssen2014}, the duration of the star-formation process, $\tau_{\rm sf}$, averaged over all the individual star-forming regions within the breakdown-scale aperture, $\Delta x_{\rm b}$, can be expressed as 
   \begin{equation}
      \label{eq:tau_sf}
      \tau_{\rm sf} = \Delta t_{\rm min} \left( \frac{\Delta x_{\rm b}}{\lambda_{\rm sep}} \right)^2\ {\rm Myr}\,,
   \end{equation}
   where $\Delta t_{\rm min}$ is the duration of the shortest phase of the star-formation process, and $\lambda_{\rm sep}$ is the characteristic spatial separation of independent star-forming regions. We assume that each individual star-forming region in SMM\,J0658 is diphasic, consisting of a molecular gas and a stellar phase. The total duration of the star-formation process is then given by $\tau_{\rm sf}=t_{\rm CO}+t_{\rm star}-t_{\rm over}$, where $t_{\rm CO}$ is the molecular gas phase duration, $t_{\rm star}$ is the duration of the young stellar phase, and $t_{\rm over}$ is the overlap phase during which young stars and molecular gas coexist \citep{Kruijssen2014, Kruijssen2018}. We adopt $\Delta t_{\rm min}$\,=\,$t_{\rm star}$\,=\,5~Myr for the young stellar populations \citep{Kennicutt2012}, and assume $t_{\rm over}$\,=\,1.5~Myr, corresponding to the stellar-feedback timescale for cloud dispersal \citep[e.g.,][]{Kruijssen2018}. In disk-like systems, a proxy for $\lambda_{\rm sep}$ is the Toomre length, $l_{\rm T}$ \citep{Escala2008}, expressed as 
   \begin{equation}
      \label{eq:toomre_l}
      l_{\rm T} =  \frac{\pi \, G \, \Sigma_{\rm disk}}{\Omega^2}\ {\rm kpc}\,,
   \end{equation}
   where $\Sigma_{\rm disk}$ is the disk gas mass surface density, and $\Omega$\,=\,$V/R_{1/2}$ is the angular velocity at circular velocity $V$. 
   We measured $\Sigma_{\rm disk}$ in the 0\farcs2 source-plane maps after masking all resolved CO and dust clumps, and verified that this value does not vary significantly when the unresolved clumps are also masked. We find $l_{\rm T}$\,=\,$1.1\pm0.3$~kpc, about five times larger than typical values in local galaxies studied down to $\sim$50~pc \citep[e.g.,][]{Chevance2020}. However, this separation scale is consistent with the effective clump diameters of $\sim$460\,--\,960~pc in the 0\farcs2 ALMA maps (see Table~\ref{tab:GMC_prop}). We note that $\lambda_{\rm sep}$\,$>$\,$\Delta x_{\rm b}$ implies $\tau_{\rm sf}$\,$<$\,$t_{\rm min}$, leading to unphysical, negative GMC lifetimes. 
   However, local studies have shown that $\lambda_{\rm sep}$ is often much smaller than the breakdown scale, with mean separations of 100\,--\,300~pc in local star-forming disks \citep[][]{Chevance2020}. 
   Assuming that the observed spatially resolved clumps in the 0\farcs2 ALMA maps are clusters of smaller CO and dust clumps, consistent with the morphological differences between the 0\farcs6 and 0\farcs2 maps (see Fig.~\ref{fig:KS_GR26}), we expect smaller separation scales between independent star-forming regions. Under this assumption, if we instead adopt $\lambda_{\rm sep}$\,=\,300~pc, corresponding to the scale at which a plateau is reached in each branch of the fork diagram (see Fig.~\ref{fig:KS_fork}) and consistent with $\lambda_{\rm sep}$ values being generally inferred from this region of the fork \citep[e.g.,][]{Chevance2020, Ramambason2026}, we derive a GMC lifetime of $\sim$20~Myr. If we adopt $\lambda_{\rm sep}$\,=\,200~pc, consistent with values for submillimeter galaxies at 1\,$<$\,$z$\,$<$\,2.5 \citep{Genzel2010}, we derive a GMC lifetime of $\sim$60~Myr. The resulting range of $t_{\rm CO}$\,$\sim$20\,--\,60~Myr is about a factor of two longer than the typical GMC lifetimes of $\sim$10\,--\,30~Myr measured in local star-forming galaxies \citep[][]{Chevance2020}. 
   
   \citet{Kruijssen2018} have shown that GMCs located in the inner parts of galaxies have longer lifetimes, likely because higher disk stability, possibly linked to morphological quenching, supports them against gravitational collapse. We note that the ALMA observations mainly capture emission from the inner part of SMM\,J0658, while the outskirts are too faint (see Fig.~\ref{fig:jwst}). Additionally, the high molecular gas fractions observed at $z$\,$>$\,2 may imply larger molecular gas reservoirs \citep{Forster2020}, which may require longer timescales to be dispersed than smaller local GMCs. However, $\Delta x_{\rm b}$ and $\lambda_{\rm sep}$ are degenerate quantities that depend strongly on the observing angular resolution, and our GMC lifetime estimates should therefore be interpreted with caution. Our measurements may be biased by a combination of factors, including (i)~uncertainties in the physical quantities used in this study, (ii)~the stellar tracer used, (iii)~the high disk inclination, and (iv)~the angular resolution limit of the ALMA data, likely leading to significant clump blending. Further spatially resolved studies are required to test these possibilities and better constrain how GMC lifetimes depend on the ISM and disk conditions in SMM\,J0658.

\subsection{Lens model dependence and possible caveats}
   \label{sec:disc_caveats}
   Compared to previous work using other lens models \citep[][]{Gonzalez2010, Johansson2012, Motta2018}, the molecular and stellar masses of SMM\,J0658 are revised upward, consistent with the lower magnification predicted by the JWST-based lens model, while the SFR is revised downward using new photometric data. The ray-traced approach used in this work does not account for lensing point-spread-function effects, which may affect the reconstructed morphology and sizes of resolved structures. Forward modeling methods that address this \citep[e.g.,][]{Sharma2018, Sharma2021} are beyond the scope of the present study. We note here that the lens model uncertainty is smaller than the one associated with the physical quantities. Moreover, surface densities are not directly affected by lensing uncertainties, since they are computed as ratios, i.e., magnification cancels out. 
   
   Several physical assumptions affect $\Sigma_{\rm mol}$ and $\Sigma_{\rm SFR}$. These assumptions were adopted at the galaxy scale, while our analysis relies on spatially resolved maps. Notably, simulations have shown that the $\alpha_{\rm CO}$ conversion factor can increase significantly in the galaxy central parts due to CO depletion \citep[e.g.,][]{Narayanan2012, Anirudh2025}. The main observational bias is the high disk inclination of SMM\,J0658, which implies that the ALMA data may reveal structures lying along the same line of sight. Still, the lensing configuration of SMM\,J0658 allows us to probe the ISM across the disk at sub-kpc scales, which have rarely been reached at $z$\,$>$\,2 before \citep[e.g.,][at $z$\,$\sim$\,1]{Nagy2023}. 
   The largest uncertainty is related to the derivation of the CO-to-H$_2$ conversion factor. Although ATCA \text{CO(1--0)} and \text{CO(3--2)} observations remove the line luminosity ratio-based uncertainty in deriving $M_{\rm mol}$, $\alpha_{\rm CO}$ value can still vary by an order of magnitude across different ISM environments \citep[e.g.,][]{Bolatto2013, Tacconi2013, Sun2023}. In addition, CO observations may miss a CO-dark molecular gas component, leading to an underestimate of the total molecular gas content \citep[e.g.,][]{Wolfire2010, Madden2020}. 
   The $\Sigma_{\rm SFR}$ is constrained from the galaxy SED modeling, including NIRCam photometry and ALMA flux densities and therefore accounting for the obscured star formation. However, SED-based estimates remain sensitive to several assumptions, such as the adopted star-formation history and stellar metallicity \citep[][]{Boquien2019}. 
   
   The range of GMC lifetime estimates from the tuning-fork diagram should be interpreted with caution. Face-on galaxies are ideal systems to probe the spatial decorrelation between molecular gas and young stars, whereas SMM\,J0658 is a highly inclined disk-like system. Projection effects likely blend distinct regions and may bias the apparent separation between molecular gas- and dust-dominated clumps. In addition, the rest-frame 803~$\upmu$m dust continuum traces a wider range of stellar ages than H$\alpha$ \citep[e.g.,][]{Kennicutt2012, Leroy2013}. Ideally, reproducing such local-galaxy studies at $z$\,$>$\,2 would require using sub-kpc H$\alpha$ maps, e.g., from JWST/NIRSpec IFU observations, together with stellar and cold-gas tracers resolved down to physical scales of $\sim$50~pc. This remains challenging even for magnified arcs, because strong lensing conserves the source surface brightness and achieving higher angular resolutions require substantially longer integration times. The ALMA Wideband Sensitivity Upgrade \citep{Carpenter2023}, with its improved continuum and spectral line imaging sensitivity, may make such cold-gas studies feasible in the future by enabling deeper observations of faint, extended emission at very high angular resolution.

\section{Conclusions}
\label{sec:conclusions}
   In this study, we used spatially resolved ALMA observations of the \text{CO(3--2)} and rest-frame 803\,$\upmu$m dust-continuum emission in SMM\,J0658 to test the KS relation on sub-kpc scales at $z$\,$\sim$\,$2.78$. Using a JWST-based lens model, we reconstructed source-plane maps of $\Sigma_{\rm mol}$, $\Sigma_{\rm SFR}$, and $\tau_{\rm dep}$ down to physical scales of $\sim$200~pc. This allowed us to investigate the depletion time scale dependence at $z$\,$>$\,2. 
   Our main results can be summarized as follows:
      \begin{itemize}
         {\smallskip}
         \item[$\bullet$] The 0\farcs2 ALMA data reveal dense star-forming clumps with $\Sigma_{\rm mol}$\,$\sim$\,1.3--2.3~$\times10^3~{\rm M}_\odot\,{\rm pc}^{-2}$, $\Sigma_{\rm SFR}$\,$\sim$\,0.5--2.3~${\rm M}_\odot\,{\rm yr}^{-1}\,{\rm kpc}^{-2}$, and $\tau_{\rm dep}$\,$\sim$\,0.82\,--\,2.56~Gyr. At the current global SFR, the molecular gas reservoir of SMM\,J0658 would be exhausted within $\tau_{\rm dep}$\,$\leq$\,1.06\,$\pm$\,0.27~Gyr assuming continuous star formation, consistent with SMM\,J0658 lying on the main sequence at $z$\,$\sim$\,$2.78$. The observed clumpy distributions of molecular gas and dust-continuum emission are consistent with localized star-forming regions within the ISM. 
         {\smallskip}
         \item[$\bullet$] The 0\farcs2 and 0\farcs6 ALMA data reveal a clear CO--dust spatial decorrelation, indicating localized molecular gas reservoirs and dust-obscured star-forming regions across the disk. We quantify the scale dependence of the depletion time across the source by measuring CO-to-dust flux ratios over apertures ranging from 200~pc to 3.2~kpc, and find a breakdown scale of 0.8\,$\pm$\,0.1~kpc in SMM\,J0658 at 0\farcs2 angular resolution. Our results are consistent with a scale-dependent breakdown of the KS relation driven by the resolution of individual evolutionary stages of star formation in the ISM of SMM\,J0658, in line with observations from the local universe to $z$\,$\sim$\,1. We extend this picture to $z$\,$\sim$\,$2.78$ using dust continuum as a tracer of active star-forming regions. 
         {\smallskip}
         \item[$\bullet$] We find evidence that SMM\,J0658 and SMG-2 are not multiple images of the same background galaxy, but are instead two distinct systems at $z$\,$\sim$\,$2.78$. The model-predicted physical separation of 60\,$\pm$\,5~kpc in the source plane and the measured rest-frame recessional velocity offset of $\sim$120~${\rm km\,s^{-1}}$ are consistent with a galaxy pair and merger candidate. Both sources appear morphologically undisturbed in NIRCam imaging, with no clear tidal features, suggesting either no ongoing interaction or an early interaction stage. 
      \end{itemize}

   Future ALMA observations may help us expand this unique high-$z$ clump sample to galaxies spanning diverse ISM conditions at $1.5$\,$\lesssim$\,$z$\,$\lesssim$\,$6$ (PIDs 2026.1.00005.S and 2026.1.00471.S, PI~Cornil-Ba\"{i}otto), for testing whether the properties of the star-forming complexes identified in SMM\,J0658 are representative of the cosmic-noon population.
   Our tentative GMC lifetime estimate, $t_{\rm CO}$\,$\sim$20--60~Myr at $z$\,$\sim$\,2.78, is about a factor of two longer than typical local GMC lifetimes, but may be affected by our physical assumptions and observational biases, and therefore requires further investigation. It would be valuable to compare this result with maps of young-stellar tracers, such as H$\alpha$ from JWST/NIRSpec or Pa$\alpha$ from MIRI/MRS IFU observations.
   This work highlights the power of combining high-resolution ALMA observations with strong gravitational lensing to probe how star formation proceeds on sub-kpc scales in main-sequence galaxies at cosmic noon, a key epoch for galaxy evolution.

\begin{acknowledgements}
   We thank the anonymous referee for the thorough feedback and helpful comments.
   C.C. thanks Miroslava Dessauges-Zavadsky for valuable discussions during the European Astronomical Society (EAS) Annual Meeting 2026. 
   This research makes use of the following ALMA data: ADS/JAO.ALMA \#2015.1.01559.S, \#2018.1.01754.S, and \#2024.1.00015.S. ALMA is a partnership of ESO (representing its member states), NSF (USA) and NINS (Japan), together with NRC (Canada), MOST and ASIAA (Taiwan), and KASI (Republic of Korea), in cooperation with the Republic of Chile. The Joint ALMA Observatory is operated by ESO, AUI/NRAO and NAOJ. 
   This work is also based on observations made with the NASA/ESA/CSA James Webb Space Telescope. 
   These observations are associated with the program JWST-GO-04586. We thank the GO 4598 team for sharing the reduced NIRCam images and photometric catalogs of the Bullet Cluster.
   C.C. acknowledges support from Agencia Nacional de Desarrollo y Investigación (ANID) through its Scholarship Program, Doctorado Becas Chile No.~21250625, and partial support from FONDECYT through grant No.~1221846. 
   J.M. and E.I. gratefully acknowledges support from ANID MILENIO NCN2024\_112. 
   G.R. and M.B. acknowledge support from the ESA PRODEX Experiment Arrangement No. 4000146646, ERC Grant FIRSTLIGHT, and Slovenian National Research Agency ARIS, through grants N1-0238 and P1-0188. 
   M.B. acknowledges support by the ANID BASAL project FB210003. This work was supported in part by the French government through the France 2030 investment plan managed by the National Research Agency (ANR), as part of the Initiative of Excellence of Universit\'{e} C\^{o}te d'Azur under reference No.~ANR-15-IDEX-01. This research was funded in part by the French National Research Agency (ANR), grant ANR-25-CE31-0192 (project CIGALE2D). 
   T.V. acknowledges support from the Secretar\'{i}a de Ciencia, Humanidades, Tecnolog\'{i}a e Innovaci\'{o}n (SECIHTI) under grant CBF-2025-I-551 (“Convocatoria Ciencia B\'{a}sica y de Frontera 2025”). 
   This research also made use of \texttt{Astropy}\,\footnote{\url{https://www.astropy.org}}, a community-developed core \texttt{Python} package for Astronomy \citep{astropy2022}.
   
\end{acknowledgements}

\section*{Data availability}
   This work made use of ALMA observations that can be retrieved from the ALMA Science Archive\footnote{\url{https://almascience.nrao.edu/aq/}} under PIDs~2015.1.01559.S, 2018.1.01754.S, and 2024.1.00015.S. 
   The lens model used in this analysis was developed by \citet{Rihtarsic2026} and can be retrieved from the CAnadian NIRISS Unbiased Cluster Survey\footnote{\url{https://niriss.github.io/lensing.html}} (CANUCS), as well as in the Strong Lensing Cluster Atlas Data Base\footnote{\url{https://data.lam.fr/sl-cluster-atlas/home}} hosted at the Laboratoire d'Astrophysique de Marseille. \\

\bibliographystyle{aa}
\bibliography{references}{}

\appendix
\onecolumn

\section{0\farcs6 ALMA data}
\label{app:B3_IP}
\begin{figure*}[ht]
    \centering
	\includegraphics[width=0.99\textwidth]{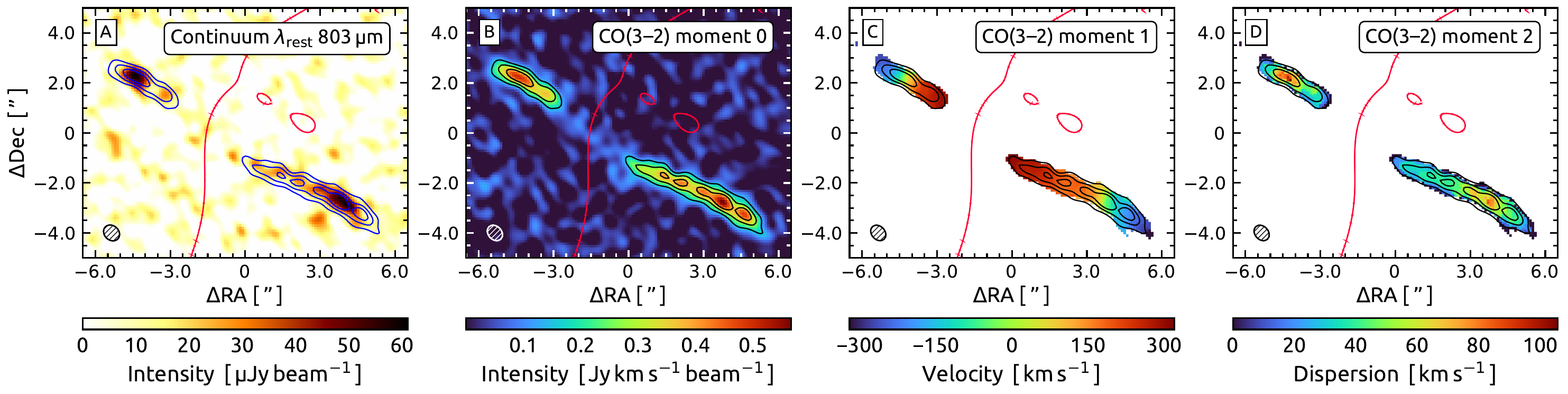}
    \caption{
    {\bf A:}~Rest-frame 803\,$\upmu$m continuum map at 0\farcs6. 
    {\bf B:}~\text{CO(3--2)} velocity-integrated flux map at 0\farcs6 and $25.6\,{\rm km}\,{\rm s}^{-1}$ spectral resolution.  
    {\bf C:}~\text{CO(3--2)} rotation velocity map, and 
    {\bf D:}~\text{CO(3--2)} velocity dispersion map, both with a 5$\sigma$ threshold.
    In each map, \text{CO(3--2)} contours at 4$\sigma$, 7$\sigma$, and 10$\sigma$-significance are shown. Each ellipse is the synthesized beam. The critical lines at $z_{\rm CO}$ from the \citetalias{Rihtarsic2026} lens model are plotted in red. 
    }
    \label{fig:B3_IP}
\end{figure*}

    Figure~\ref{fig:B3_IP} presents the 0\farcs6 ALMA Band-3 data covering the lensed images A and B of SMM\,J0658. The mirror symmetry between A and B in the rest-frame 803\,$\upmu$m continuum and \text{CO(3--2)} moment maps, most clearly seen in the reversed CO velocity field, is due to the parity flip induced by strong gravitational lensing. The recessing side of the velocity field is also amplified by strong lensing. 

\section{Magnification map}
\label{app:mu_map}
\begin{figure*}[ht]
    \centering
    \includegraphics[width=0.90\textwidth]{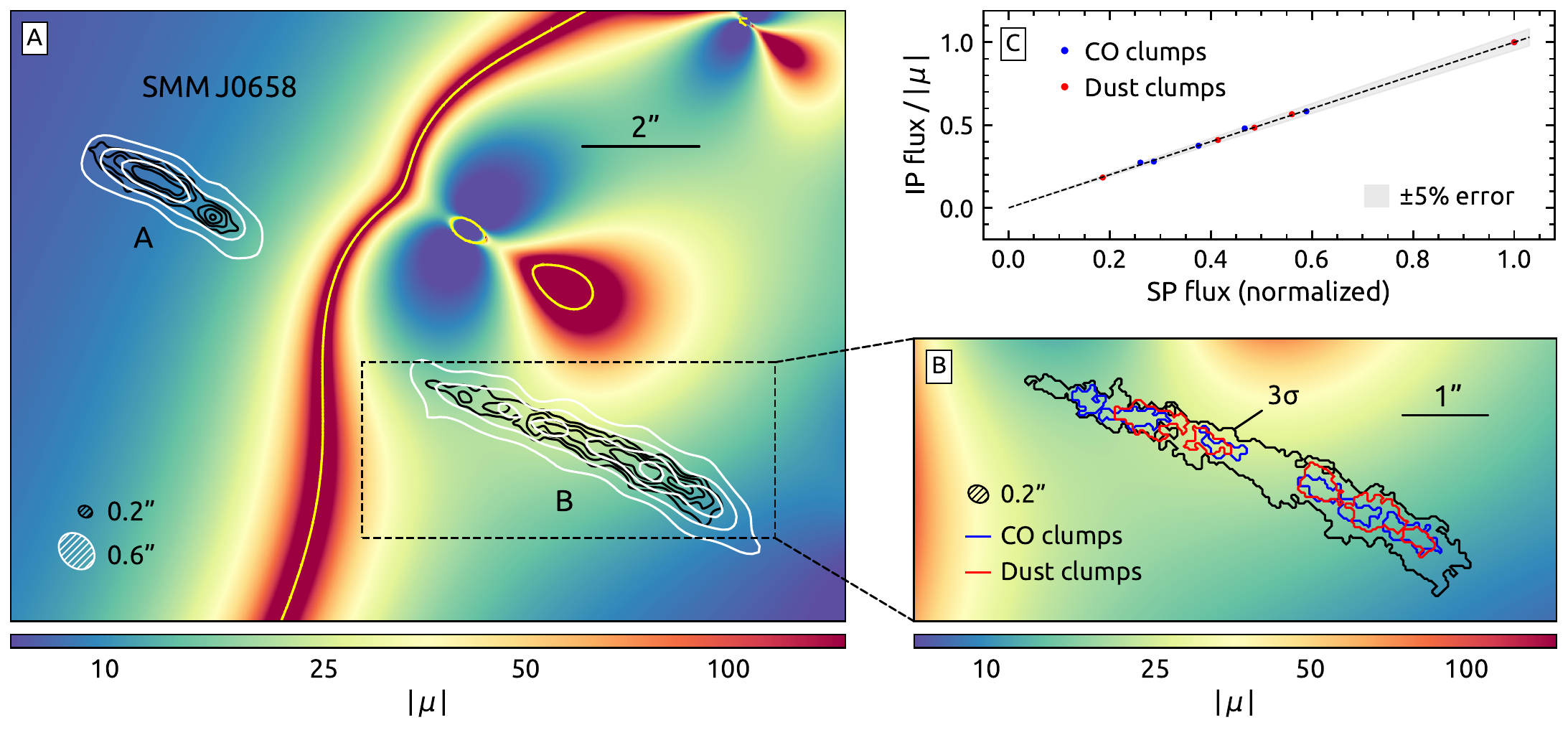}
    \caption{ 
    {\bf A:}~Absolute magnification map generated with the \citetalias{Rihtarsic2026} lens model. The model-predicted critical lines at $z_{\rm CO}$\,=\,2.7768 are plotted in yellow. ALMA contours of the \text{CO(3--2)} emission at 0\farcs2 (0\farcs6) resolution are shown in black (white) at $3\sigma$, $4\sigma$, $5\sigma$, and $6\sigma$ ($3\sigma$, $6\sigma$, $9\sigma$, and $12\sigma$) levels. The ellipses are the corresponding ALMA synthesized beams. 
    {\bf B:}~Zoom-in view of the magnification map covering Image~B of SMM\,J0658. We show the spatial locations of the resolved clumps of \text{CO(3--2)} and rest-frame 803\,$\upmu$m dust-continuum at 0\farcs2 resolution. \text{CO(3--2)} contours at $3\sigma$ significance (black) are overlaid. The ellipse is the synthesized beam. 
    {\bf C:}~We compared clump-integrated fluxes measured in the 0\farcs2 ALMA data in the image plane (IP), corrected for the average clump magnification, with fluxes measured directly in the ray-traced source plane (SP) of the \text{CO(3--2)} moment-0 and continuum maps. All fluxes were normalized to their peak values. Shaded regions correspond to the $5\%$ boundary. 
    }
    \label{fig:mu_map}
\end{figure*}

   Figure~\ref{fig:mu_map} presents the model-predicted absolute magnification map covering the images A and B of SMM\,J0658 at $z_{\rm CO}$\,=\,$2.7768$ (panel~A). The map is generated from the best-fit JWST-based lens model of \citetalias{Rihtarsic2026}. The mean absolute magnification over $3\sigma$ \text{CO(3--2)} emission in the 0\farcs2 (0\farcs6) ALMA data is $\langle \mu_{\rm A} \rangle=9.1^{\smash{+0.8}}_{\smash{-0.2}}$ ($9.8^{\smash{+1.0}}_{\smash{-0.2}}$) for Image~A and $\langle \mu_{\rm B} \rangle=19.2^{\smash{+2.5}}_{\smash{-0.8}}$ ($18.2^{\smash{+2.4}}_{\smash{-0.5}}$) for Image~B. 
   Since the magnification gradient across individual CO and dust clumps is small (see panel~B), clump-integrated flux measurements performed directly in the source plane, where the magnification is corrected pixel by pixel, agree well with image-plane measurements corrected using the average magnification over each clump region (see panel~C). 

\newpage

\section{Source-plane reconstructed beams}
\label{app:SP_beams}
\begin{figure*}[ht]
    \centering
	\includegraphics[width=0.99\textwidth]{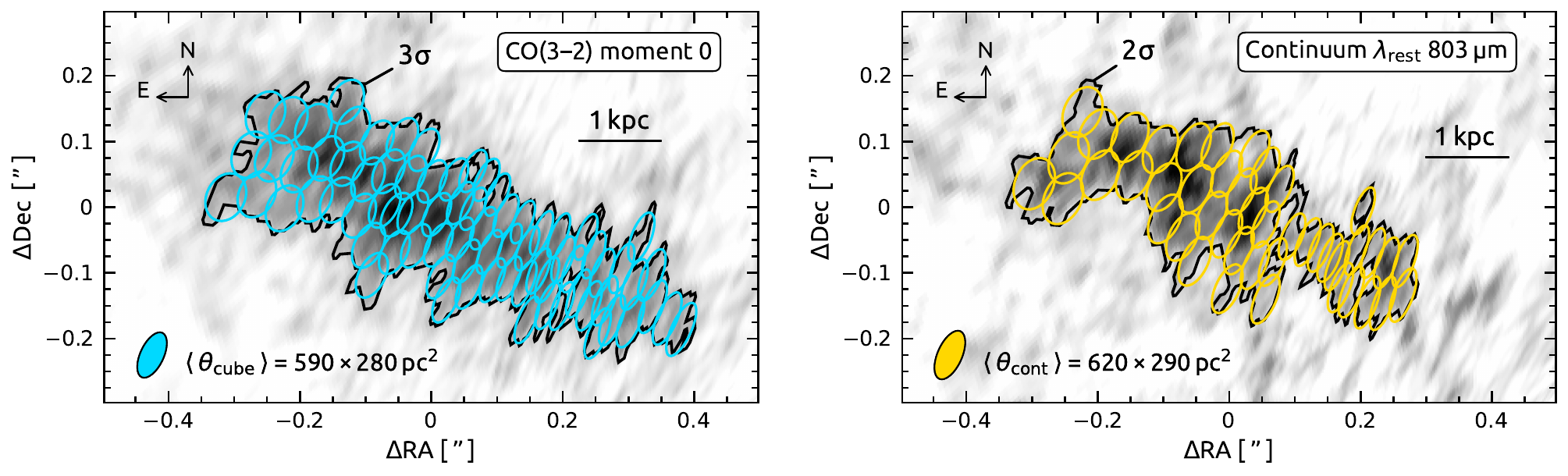}
    \caption{
    Spatial variation of the source-plane reconstructed 0\farcs2 synthesized beams in SMM\,J0658 at $z_{\rm CO}$\,=\,2.7768, obtained by reconstructing Image~B with the \citetalias{Rihtarsic2026} lens model. 
    {\bf Left:}~Source-plane \text{CO(3--2)} velocity-integrated flux map, overlaid with \text{CO(3--2)} contours at 3$\sigma$-significance in black, and the ray-traced beams in blue. The averaged beam is shown as a filled ellipse. 
    {\bf Right:}~Source-plane rest-frame 803\,$\upmu$m continuum map, overlaid with continuum contours at 2$\sigma$-significance in black, and the ray-traced beams in yellow.
    }
    \label{fig:SP_beams}
\end{figure*}

     In the ALMA observations (i.e., image-plane maps), the angular resolution is uniform since the synthesized beam is constant. In the source-plane maps, however, the physical scale is uniform, while the beam varies across the reconstructed source. 
     Figure~\ref{fig:SP_beams} shows the spatial distribution of the beams in the source plane. The beams are not identical in the \text{CO(3--2)} and dust continuum maps for two reasons: (i)~CO and continuum emission do not have the same morphology, so the beam is not ray-traced at exactly the same location in the two maps; (ii)~independently from the imaging weights, the beam size differs slightly between the CO and continuum data because of their different observing frequencies. Specifically, the spectral line spans a limited number of velocity channels ($\nu_{\rm CO}$\,=\,$91.558$~GHz), whereas the continuum is averaged over the full bandwidth of the four spectral windows ($\nu_{\rm cont}$\,$\approx$\,98.8~GHz).

\section{Source-plane reconstructed clumps}
\label{app:SP_dendro}
\begin{figure*}[ht]
    \centering
	\includegraphics[width=0.99\textwidth]{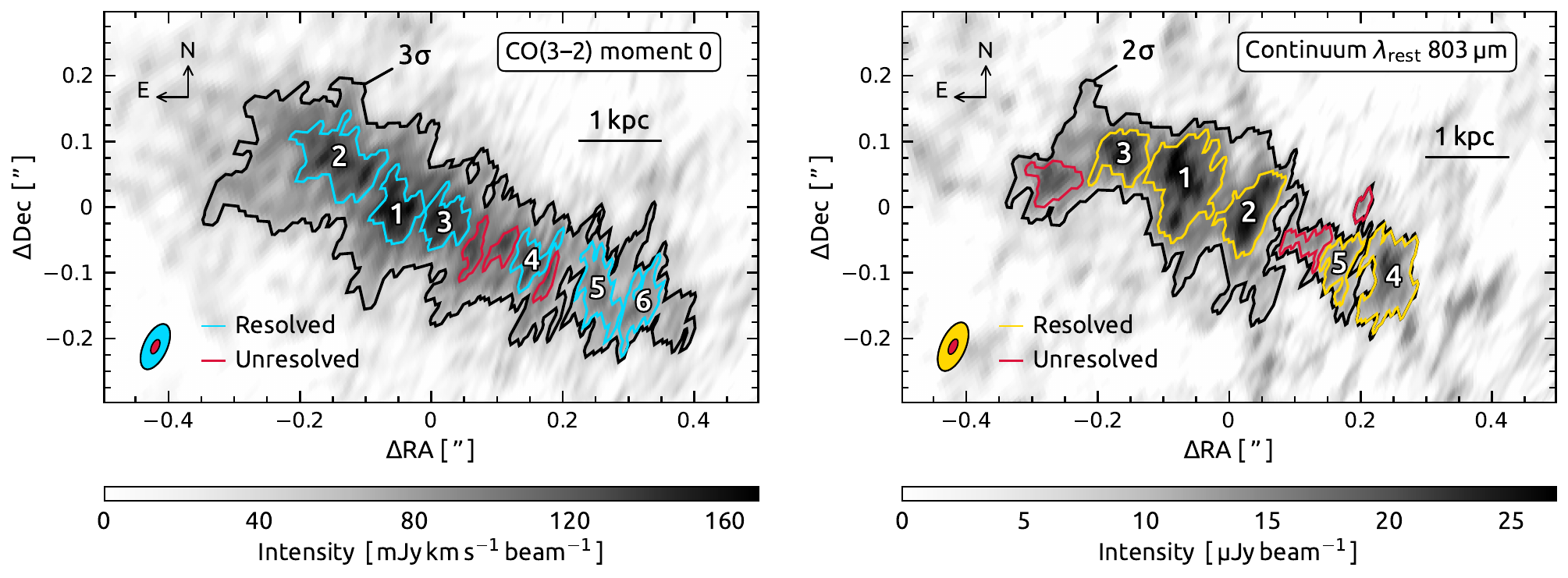}
    \caption{
    Ray-traced CO and dust clumps at $z_{\rm CO}$, reconstructed using the \citetalias{Rihtarsic2026} lens model.
    {\bf Left:}~Source-plane \text{CO(3--2)} velocity-integrated flux map, overlaid with \text{CO(3--2)} contours at 3$\sigma$ significance. The averaged reconstructed 0\farcs2 (0\farcs02) synthesized beam corresponds to the blue (red) ellipse. 
    {\bf Right:}~Source-plane rest-frame 803\,$\upmu$m continuum map, overlaid with 2$\sigma$ dust-continuum contours. The averaged reconstructed 0\farcs2 (0\farcs02) beam corresponds to the yellow (red) ellipse. 
    The properties of the spatially resolved clumps are listed in Table~\ref{tab:GMC_prop}.
    }
    \label{fig:SP_dendro}
\end{figure*}

    We used custom {\small PYLENSTOOL} routines following the source-reconstruction methods implemented in {\small LENSTOOL} to recover the intrinsic fluxes, spatial locations, and morphologies of the clumps identified in Image~B of SMM\,J0658 with {\small ASTRODENDRO}. Figure~\ref{fig:SP_dendro} shows the ray-traced source-plane locations of the \text{CO(3--2)} and dust-continuum clumps at $z_{\rm CO}$\,=\,$2.7768$, using the JWST-based lens model of \citetalias{Rihtarsic2026}. The corresponding flux dendrograms are shown in Figure~\ref{fig:dendro}. The clump positions are reversed relative to Figure~\ref{fig:dendro} due to the parity flip induced by strong gravitational lensing.

\newpage

\section{Galaxy spectral energy distribution}
\label{app:SED}
\begin{figure*}[ht]
    \centering
	\includegraphics[width=0.99\textwidth]{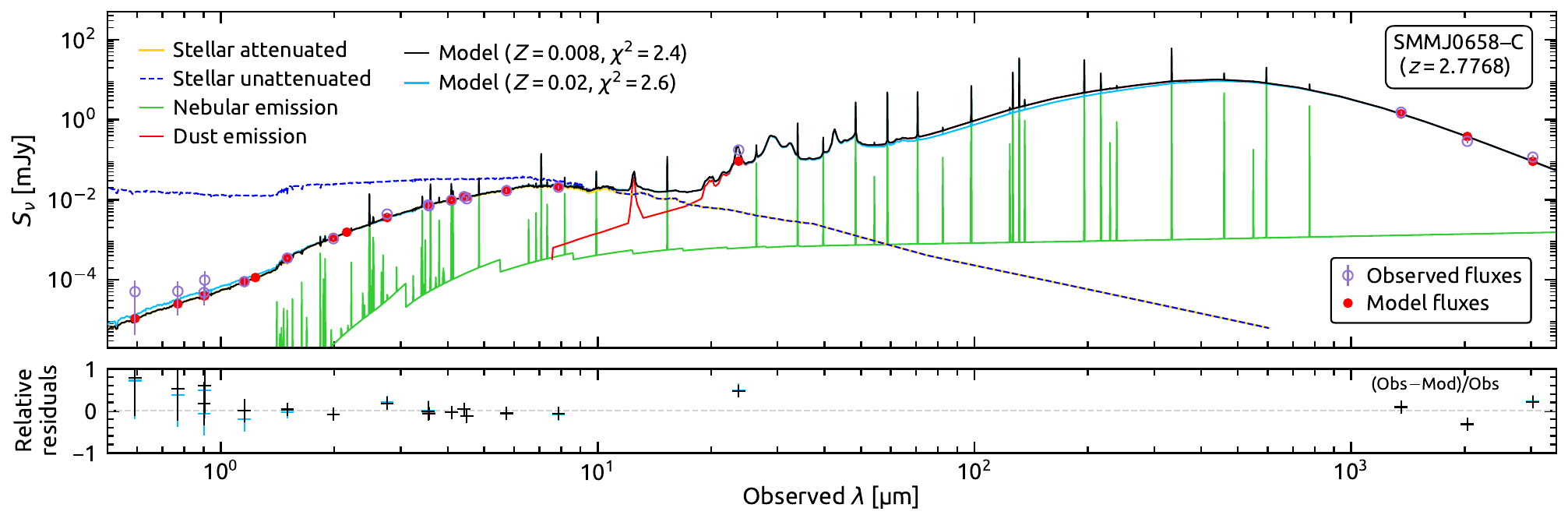}
    \caption{
    Best-fit SED models of Image~C of SMM\,J0658 at $z_{\rm CO}$\,=\,$2.7768$ using {\small CIGALE}. The black solid curve shows the best model spectrum obtained for a stellar metallicity of $Z$\,=\,$0.008$ ($\chi^2$\,=\,$2.4$), while the blue curve shows the solar-metallicity model ($\chi^2$\,$=$\,$2.6$). The individual model components (stellar, nebular, and dust emission) correspond to the $Z$\,$=$\,$0.008$ model. The lower panel shows the relative residuals for both models. 
    }
    \label{fig:sed}
\end{figure*}

   Figure~\ref{fig:sed} presents the SED modeling of SMM\,J0658 over the rest-frame $\sim$0.5--800~$\upmu$m using {\small CIGALE}\footnote{\url{https://gitlab.lam.fr/cigale/cigale/}} \citep{Boquien2019}. Compared to previous SED analyses \citep[e.g.,][]{Gonzalez2010, Johansson2012}, our models include NIRCam photometry and ALMA continuum flux densities in Bands~3, 4, and 6 (PIDs 2015.01559.S and 2024.1.00015.S). We used photometric catalogs from the CANUCS Data Releases\footnote{\url{https://niriss.github.io/data.html}}, including HST and JWST data, together with \textit{Spitzer} photometry from \citet{Gonzalez2010}, and discarded the 2MASS $J$ and $K_{\rm S}$ upper-limit measurements. 
   We used the flux densities of Image~C for two reasons: (i)~they are less affected by contamination from foreground objects; and (ii)~Images~A and B merge along the critical lines in the NIRCam images (see Fig.~\ref{fig:jwst}). Since Image~C is not covered by the 0\farcs6 ALMA observations and is not detected in the 0\farcs2 ALMA data, the ALMA fluxes were measured from Images~A and B and rescaled to the best-fit magnification of Image~C, $\langle \mu_{\rm C} \rangle$\,=\,$5.5^{\smash{+0.2}}_{\smash{-0.3}}$. The dust emission in {\small CIGALE} is modeled using a power-law distribution of dust mass over the intensity of the stellar radiation field \citep{Dale2001}. 
   We tested two stellar metallicities, $Z$\,$=$\,$0.02$ and $0.008$, and found no significant variation. For the analysis, we adopted the best-fit global estimates of $M_\ast$\,=\,$(4.0\pm1.0)\times10^{10}\ {\rm M}_\odot$, $A_V$\,=\,$3.3\pm0.8$, and SFR averaged over the last 10~Myr, ${\rm SFR}_{\smash{10\,{\rm Myr}}}$\,=\,$47\pm12\ {\rm M}_\odot\,{\rm yr}^{-1}$ from the $Z$\,$=$\,$0.008$ model, which has the lowest reduced $\chi^2$. We adopt ${\rm SFR}_{\smash{10\,{\rm Myr}}}$ because it traces recent star formation for the KS study. We also verified that these estimates are consistent with the mass--metallicity relation at $z$\,$\sim$\,$2.78$ \citep[e.g.,][]{Sanders2021}.

\section{Lens model comparison}
\label{app:sp_tau}
\begin{figure*}[ht]
    \centering
	\includegraphics[width=0.99\textwidth]{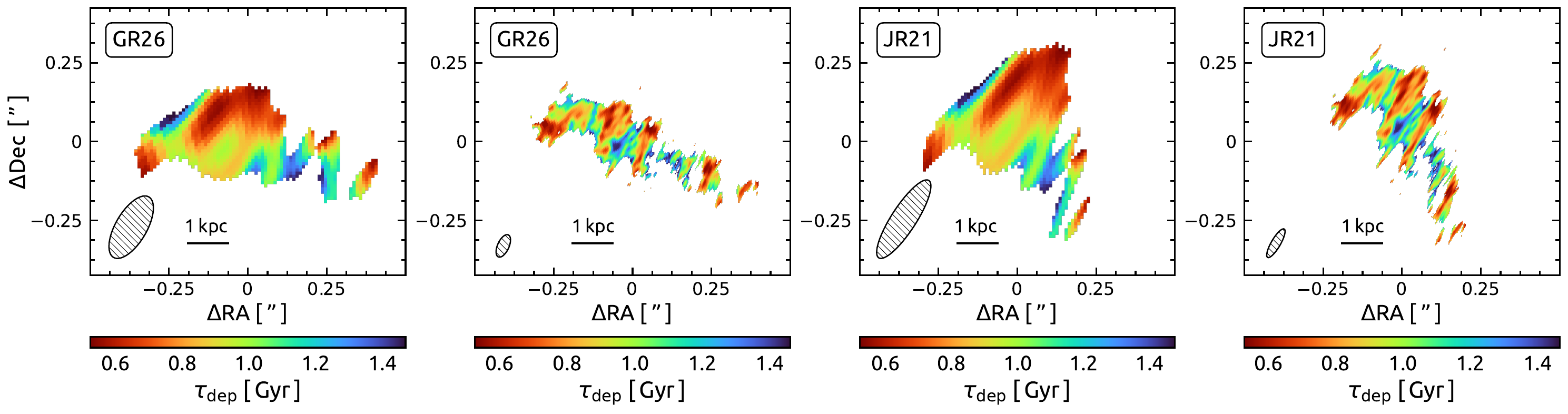}
    \caption{
    Source-plane maps of molecular gas depletion time, computed as $\tau_{\rm dep}$\,$\equiv$\,$\Sigma_{\rm mol}$/$\Sigma_{\rm SFR}$ after applying a $3\sigma$ threshold, and derived from the ALMA 0\farcs6 and 0\farcs2 data using the \citetalias{Rihtarsic2026} and \citetalias{Richard2021} lens models. In each map, the ellipse shows the average source-plane reconstructed synthesized beam. The corresponding average physical beam sizes at $z_{\rm CO}$\,=\,$2.7768$ are 1270 and 450~pc for the 0\farcs6 and 0\farcs2 \citetalias{Rihtarsic2026} reconstructions, and 1460 and 530~pc for the 0\farcs6 and 0\farcs2 \citetalias{Richard2021} reconstructions, respectively.  
    North is up, and east is left. 
    }
    \label{fig:sp_tau}
\end{figure*}

    Figure~\ref{fig:sp_tau} compares the source-plane reconstructed $\tau_{\rm dep}$ maps at 0\farcs2 and 0\farcs6 resolution, obtained with the pre-JWST \citepalias{Richard2021} and JWST-based \citepalias{Rihtarsic2026} {\small LENSTOOL} models. The stronger magnification gradient predicted across the arc by the \citetalias{Richard2021} model leads to more elongated reconstructed beams. These differences in the recovered source morphology are expected, given the larger number of lensed galaxies with spectroscopic redshifts used in \citetalias{Rihtarsic2026}, increased by a factor of four compared to previous models. 

\newpage

   \section{Angular resolution dependence of the KS diagnostics}
   \label{app:0p6_dendro}
\begin{figure*}[ht]
    \centering
    \includegraphics[width=0.9\textwidth]{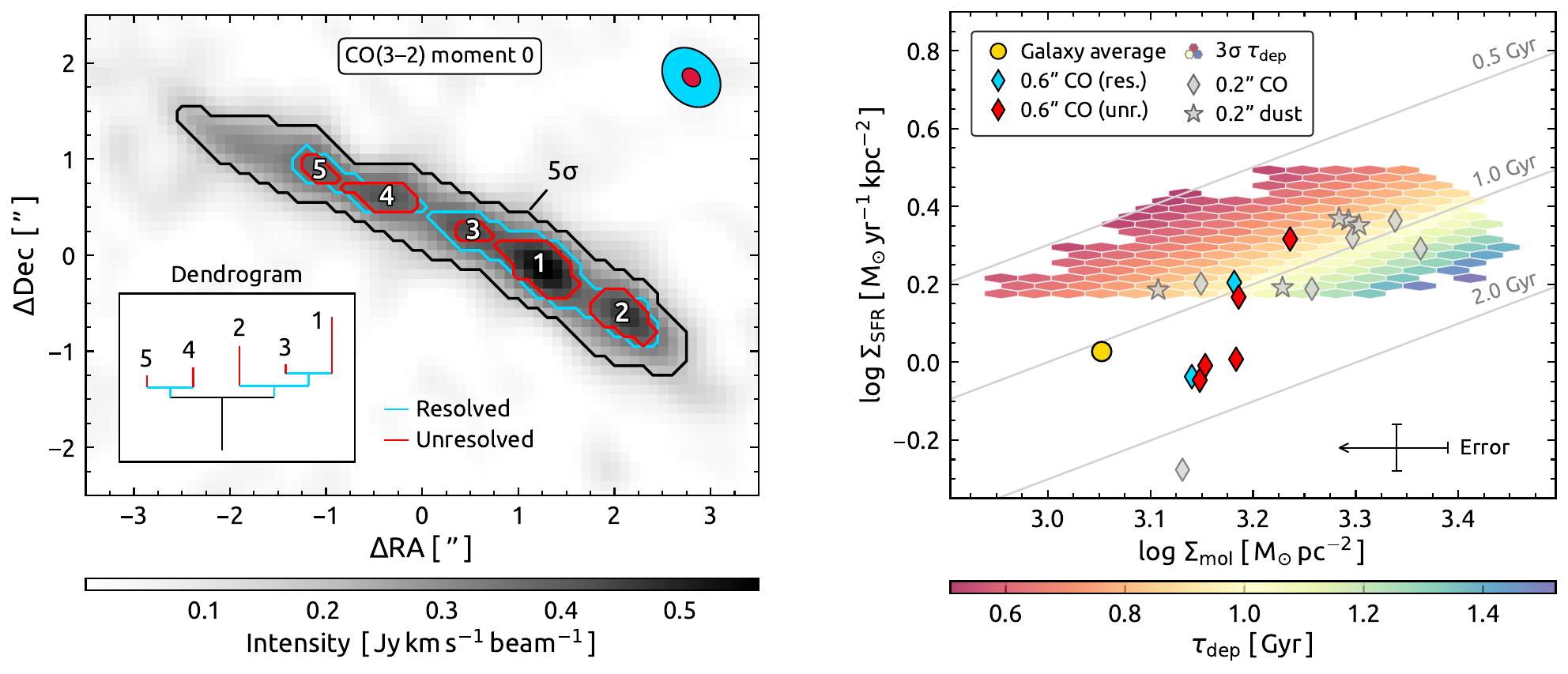}
    \caption{
    {\bf Left:}~Molecular gas complexes identified with {\small ASTRODENDRO} in the most magnified image of SMM\,J0658 (Image~B; Fig.~\ref{fig:jwst}) from the CO(3--2) velocity-integrated intensity map at 0\farcs6 resolution. The synthesized beams used for the clump extraction are shown as ellipses (0\farcs6 in cyan for the resolved structures; 0\farcs06 in red for the unresolved ones). The inset displays the corresponding dendrogram ordered by clump flux.
    {\bf Right:}~Star-forming clumps of SMM\,J0658 in the Kennicutt--Schmidt plane. The pixel-to-pixel $\Sigma_{\rm SFR}$--$\Sigma_{\rm mol}$ diagnostics for SMM\,J0658 is shown as a hexbin density map, where the background color is the molecular gas depletion time, $\tau_{\rm dep}=\Sigma_{\rm mol}/\Sigma_{\rm SFR}$, after applying a $3\sigma$ threshold to both the $\Sigma_{\rm mol}$ and $\Sigma_{\rm SFR}$ maps. Gray solid lines indicate constant depletion times of 0.5, 1.0, and 2.0~Gyr. The 0\farcs6 CO complexes (cyan) and unresolved CO substructures (red) are shown together with the resolved 0\farcs2 CO and dust-continuum clumps (gray), and the galaxy-average value (yellow).
    }
    \label{fig:0p6}
\end{figure*}

   Figure~\ref{fig:0p6} shows the molecular gas complexes identified in the 0\farcs6 ALMA CO(3--2) data using {\small ASTRODENDRO}. At this resolution, two large, kpc-scale CO complexes are resolved, while no substructures are extracted in the smoother 0\farcs6 rest-frame 803~$\upmu$m continuum emission. Although several sub-complexes are visually apparent in the 0\farcs6 CO map, they do not satisfy our clump-size criterion (iv; Sect.~\ref{sec:clump_id}), as their extracted sizes are smaller than the synthesized beam. We therefore treat these features as unresolved and exclude them from the main clump analysis. We nevertheless extracted these embedded CO sub-complexes by reducing the minimum leaf area to one-tenth of the synthesized beam (i.e., 0\farcs06), and show their positions in the KS plane together with the resolved 0\farcs2 CO and dust-continuum clumps (Fig.~\ref{fig:dendro}). Figure~\ref{fig:0p6} shows how unresolved structures affect the inferred $\Sigma_{\rm mol}$ and $\Sigma_{\rm SFR}$ to assess the impact of angular resolution on the KS diagnostics. The 0\farcs6 CO complexes distributed across the galaxy disk exhibit longer depletion times than the 0\farcs2 CO clumps. Given that the corresponding 0\farcs6 dust-continuum emission is more centrally concentrated, this may result from star-forming regions being blended toward the galaxy center at this resolution (see Sect.~\ref{sec:res_clumps}).

   Only the 0\farcs2 ALMA observations of Image~B, the most highly magnified image, allow us to measure the spatial offsets between molecular gas and star formation-dominated clumps, as shown in Figs.~\ref{fig:KS_GR26} and \ref{fig:KS_fork}. At lower angular resolution, either in Image~A or in the 0\farcs6 data, individual structures become blended into larger complexes. This blending reduces the contrast between regions dominated by molecular gas and star formation ($\lesssim$\,100~Myr stars), causing their inferred $\Sigma_{\rm mol}$ and $\Sigma_{\rm SFR}$ to converge toward the galaxy-average value \citep[e.g.,][]{Nagy2023} and reducing the spatial offsets that are central to analyzing the molecular gas depletion timescale dependence \citep[e.g.,][]{Schruba2010, Chevance2020, Ramambason2026}. Extending such detailed local studies to the early universe therefore requires spatially resolving both the cold molecular gas and star-forming regions on sub-kpc scales to distinguish the individual evolutionary phases of the star formation cycle, which becomes possible beyond the local universe by combining high-angular-resolution observations with strong gravitational lensing.

\end{document}